\documentclass[aps,11pt,prd,superscriptaddress,preprintnumbers, 
	amsfont,
	amssymb,
	amsmath,
	notitlepage,
	longbibliography,
	nofootinbib]{revtex4-1}
\usepackage[colorlinks=true, 
	linkcolor=blue, 
	citecolor=blue, 
	urlcolor=blue, 
	linktocpage=true]{hyperref}

\usepackage{graphicx,color}

\begin{document}

\title{Seeding decay of the false vacuum}
\date{\today} 

\author{Matteo Canaletti}
\email{matteo.canaletti@gmail.com}
\affiliation{School of Mathematics, Statistics and Physics, Newcastle University, 
Newcastle Upon Tyne, NE1 7RU, UK}

\author{Ian G. Moss}
\email{ian.moss@newcastle.ac.uk}
\affiliation{School of Mathematics, Statistics and Physics, Newcastle University, 
Newcastle Upon Tyne, NE1 7RU, UK}

\begin{abstract}
We present a theory of false vacuum decay induced by spherical nucleation seeds.
The type of seed considered has a boundary
characterised by surface energy terms. The theory applies to false vacuum 
decay at zero and finite temperatures. Seeded nucleation may
be important for enabling future false vacuum decay experiments on analogue
systems using Bose Einstein Condensates (BEC).  We show that our
theory of seeded nucleation at finite temperature applied to a potassium 
BEC in two spatial dimensions agrees with numerical, real-time, simulations.
\end{abstract}

\maketitle

\section{Introduction}

First order phase transitions in continuous media are characterised by the nucleation
of bubbles which grow and coalesce. A remarkable prediction of quantum field theory
is that a similar process can occur in a quantum field. 
This is the phenomenon of  false vacuum decay \cite{Coleman:1977py,Callan:1977pt}. 
False vacuum decay in elementary particle physics could have significant consequences.
In the early universe, false vacuum decay at finite temperature \cite{1983544}
could play a role in the formation of matter 
\cite{SHAPOSHNIKOV1987757} and gravitational waves \cite{10.1093/mnras/218.4.629}.
The origin of the universe could even be the result of a vacuum decay event \cite{Vilenkin:1982de}. 
The possibility of false vacuum decay in the present day was described by Coleman as 
``the ultimate ecological catastrophe" \cite{Coleman:1980aw}.
(For reviews of early universe vacuum decay see e.g. 
\cite{Mazumdar:2018dfl,10.21468/SciPostPhysLectNotes.24}.)

Familiar phase transitions in nature, for example the one between liquid and gaseous phases
of water, most often proceed with the assistance of nucleation seeds. In clouds,
dust in the upper atmosphere plays a key role in the formation of water droplets (e.g. 
\cite{https://doi.org/10.1029/2011MS000074}).
In this paper, we sketch out a general theory of nucleation near boundaries in
quantum field theory, based on ideas 
from the thermodynamical study of liquid-vapour systems (e.g \cite{Gallo2021}).

\begin{center}
\begin{figure}[ht]
\begin{center}
\scalebox{0.3}{\includegraphics{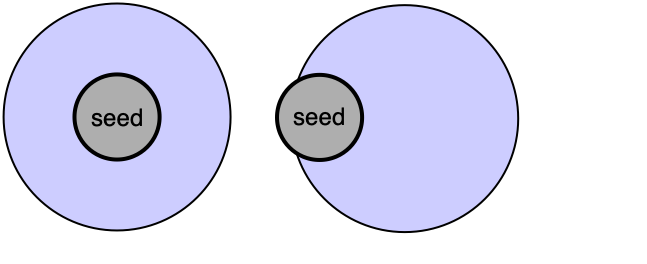}}
\end{center}
\caption{Seeded bubble nucleation. In the left figure the seed is inside a bubble of true vacuum and in the right
figure on the edge of the bubble.}
\label{fig:inter}
\end{figure}
\end{center}

For our discussion, a seed is any region which has a boundary with the true or false vacuum phase,
but is not itself in either phase. We do not consider any energy input, as might occur
in a particle collision event. We will also restrict ourselves to spherical seeds. Surface energy contributions
play an important role in the theory, and so we assume that the bubbles have sufficiently
well-defined boundaries for this.

Examples of seeded nucleation in quantum field theory exist already in the literature. 
Nucleation seeded by black holes was introduced in Refs.
\cite{PhysRevD.32.1333,Gregory_2014,Burda_2015,Burda_2015B,Burda_2016}.
Nucleation by quantum vortices was the examined in Ref. \cite{Billam:2018pvp}.
Bubbles can act like seeds in models with multiple false vacua, producing ``barnacles"
on bubble walls \cite{Czech_2012}.
Enhancement of vacuum decay due to particle collisions has also been considered 
in the past,
\cite{PhysRevD.20.3168,Selivanov:1985vt,Ellis:1990bv,Enqvist:1997wv},
although this is arguably better classed as ``catalysed" rather than seeded.

Interesting questions surround the geometry of the seeded bubble nucleation event. There are two
possibilities, the bubble surrounds the seed, which we call interstitial nucleation, or the bubble
intersects the seed, which we call edge nucleation. We show that edge nucleation
is the far more likely of the two. (This has already been seen in liquid droplet systems \cite{doi:10.1021/la104628q}.)

An important motivation for this paper has been the recent interest in experiments on false vacuum decay 
in the laboratory.  Bose-Einstein Condensates (BEC) are especially relevant because they can be 
described by a coherent quantum field.  It is possible for a multi-component BEC to have regimes
in which it behaves like a analogue system with a false vacuum state.
The first experimental demonstration of false vacuum decay at finite temperature 
has been achieved in a sodium BEC \cite{Zenesini:2023afv}.
Earlier theoretical proposals have shown that the effective theory in some systems can possess Lorentz
invariance  \cite{FialkoFate2015,FialkoUniverse2017,Billam:2022ykl}. 
In these systems, a metastable  false vacuum state exists, which is is expected to decay by
Coleman-type bubble nucleation
 \cite{Braden:2018tky,BillamSimulating2019,Jenkins:2023eez}. Simulations at finite temperature
indicate that the decay can be described by thermal instantons 
\cite{Billam:2020xna,Billam:2021psh,Pirvu:2023plk}.

In future laboratory experiments, it may useful to force bubble nucleation events in order to
study how they grow and interact. This could be done by imprinting a region with the true vacuum 
state artificially. The alternative would be to seed bubbles, for example by depleting
the density of atoms in a small region. The seeding method has the advantage of testing some
aspects of bubble nucleation theory, which the imprinting method cannot.

In the final section of this paper, we describe a numerical simulation of bubble nucleation
around a seed in a potassium--39 BEC at finite temperature, based on the scheme
proposed in \cite{Jenkins:2023npg}. The seed is created by
creating a region of high potential, which forces the density to drop. Nucleation is
enhanced, as expected, and the nucleation is always of the edge type.

\section{General considerations}

\subsection{Bubble nucleation}

\begin{center}
\begin{figure}[ht]
\begin{center}
\scalebox{0.5}{\includegraphics{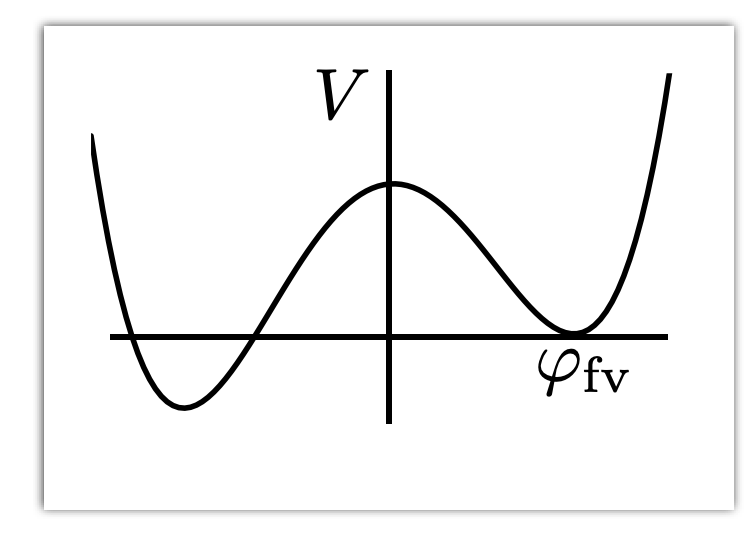}}
\end{center}
\caption{The potential of an effective field theory with true and false vacuua.}
\label{fig:inter}
\end{figure}
\end{center}

Bubble nucleation is a feature of false vacuum decay \cite{Coleman:1977py,Callan:1977pt} 
and first order phase transitions \cite{Langer:1969bc}. Both
can be described using effective field theories.  In the thermal case, the initial state is a thermal ensemble
of particle excitations about the false vacuum field $\varphi_{FV}$. We follow Ref. \cite{1983544}
in describing this case as false vacuum decay at finite temperature.
The nucleation rate of bubbles in the thermal or false vacuum state is given by a universal formula,
depending on an instanton solution $\varphi_b$ to the effective field equations. Instanton
solutions use an imaginary time coordinate, $t=i\tau$ and their action is pure imaginary $iS_E$,
where $S_E$ is termed the Euclidean action. The rate of bubble nucleation $\Gamma$
in a volume ${\cal V}$  is \cite{Coleman:1977py,Callan:1977pt} 
\begin{equation}
\Gamma={\cal V}\left|\frac{\det' S''_E[\varphi_b]}{\det S''_E[\varphi_{FV}]}\right|^{-1/2}
\left(\frac{B}{2\pi}\right)^{n/2}e^{-B},
\end{equation}
where the tunneling exponent is
\begin{equation}
B=\frac{S_E[\varphi_b]-S_E[\varphi_{FV}]}\hbar,
\end{equation}
and $S_E''$ denotes the second functional 
derivative of the action. The determinant
of an operator is the product of its eigenvalues, and a prime implies omitting $n$ zero
modes, each one of these corresponding to a translational degree of freedon
of the instanton. In the case of thermal bubble nucleation at a first order phase transition the
instantons are independent of imaginary time and the Euclidean action reduces to $S_E/\hbar=E/k_BT$,
where $E$ is the energy. (Although Planck's constant has disappeared from the exponent
in the thermal case, it can still appear in the operators, and then the tunnelling rate is set
by quantum, rather than classical, field theory.)

We will simplify the expression for the rate to
\begin{equation}
\Gamma=A_b{\cal V}B^{n/2}e^{-B}.
\end{equation}
In $D$ spatial dimensions, there are $n=D+1$ translational symmetries in the vacuum case, and $n=D$ 
in the thermal case.
The pre-factor $A_b$ is constructed from the functional determinants,
and depends on the parameters of the model.
In a relativistic system with speed of light $c$, 
a useful approximation is  that $A_b\propto R^{-D-1}c$, where $R$ is the bubble radius,
based on dimensional analysis.

Now consider what happens when the bubble nucleates around a nucleation seed, and
the number density of seeds is $n_s$.
Bubble nucleation can happen with a seed at the centre, which we call interstitial, or at the edge.
In either case, the exponent $B_s$ and the pre-factor $A_s$ will depend on the seed geometry. 
In the interstitial case, assuming some limited freedom on position of the seed inside the bubble,
and having ${\cal V}n_s$ seeds in total,
\begin{equation}
\Gamma_s=A_s{\cal V}n_sB_s^{n/2}e^{-B_s}.\label{Gamma1}
\end{equation}
In the edge case, the number of zero modes $n$ equals the surface dimension of the seed.
The zero modes will contribute a factor proportional to the surface area of the seed ${\cal A}_s$,
\begin{equation}
\Gamma_s=A_s{\cal V}n_s {\cal A}_sB_s^{n/2}e^{-B_s}.\label{Gamma2}
\end{equation}
We will make an assumption that the pre-factor $A_s$ contributes a relatively minor dependence on the seed properties.

\subsection{Phase equilibria and instantons}

The thin-wall limit is one where a bubble of a lower energy ``true" phase $TV$ is separated by a wall
from the surrounding ``false" phase $FV$, and the wall is thinner than the bubble radius $R$. 
The true phase exerts an outward pressure on the wall equal to the energy density
difference $\epsilon$ between the two phases. The size of a thermal bubble is determined by a balance between this pressure force and the
force due to surface tension of the wall $\sigma_{TF}$. In three dimensions, this is expressed by
Laplace's relation for the principle radii of curvature $R_1$ and $R_2$ of the bubble wall,
\begin{equation}
\frac{1}{R_1}+\frac{1}{R_2}=\frac{\epsilon}{\sigma_{TF}}.
\end{equation}
Laplace's relation generalises to arbitrary dimension $D$ if we use the $D-1$ principle radii
of curvature. In the fully symmetric case, the bubble radius $R$ is therefore
given by $R=(D-1)\sigma_{TF}/\epsilon$.

When any kind of barrier is introduced, there are additional surface tension terms to take into account,
\begin{itemize}
\setlength{\itemsep}{0pt}
    \setlength{\parskip}{0pt}
\item $\sigma_{ST}$ between the seed and true vacuum phase
\item $\sigma_{SF}$ between the seed and false vacuum phase
\item $\sigma_{TF}$ between true and false vacuum phases
\end{itemize}
The balance of forces at the triple boundary determines an angle of contact $\theta$ 
by Young's equation (see figure \ref{fig:compare}),
\begin{equation}
\cos\theta=\frac{\sigma_{SF}-\sigma_{ST}}{\sigma_{FT}}.\label{young}
\end{equation}
Small angles are associated with a ``wettable" surfaces, and angles close to $\pi$ with ``hydrophobic"
surfaces.

\begin{center}
\begin{figure}[ht]
\begin{center}
\scalebox{0.4}{\includegraphics{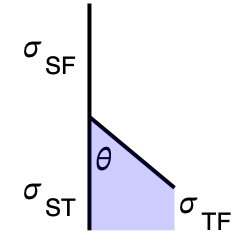}}\hfil
\scalebox{0.25}{\includegraphics{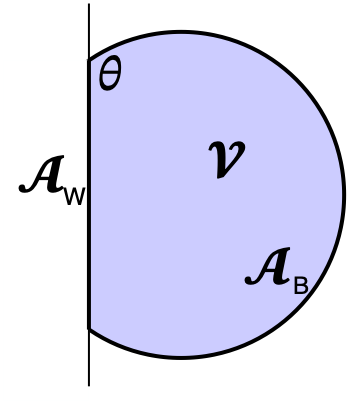}}
\end{center}
\caption{Left: Surface tensions and  angle of contact $\theta$ for the true vacuum region (shaded). 
Right: A true vacuum bubble of volume ${\cal V}$ nucleating beside a boundary, with 
contact area ${\cal A}_W$ between 
the true vacuum phase and the wall.
}
\label{fig:walls}
\end{figure}
\end{center}

An analogous situation exists for the instanton solutions in the theory of false vacuum decay at zero
temperature. An extra dimension is added to represent imaginary time and the energy is replaced by the 
Euclidean action. In a Lorentz invariant theory, there is an equivalence between dynamics
in $D$ spatial dimensions and $D+1$ Euclidean dimensions, implying that Laplace's and young's relations still hold.
(This can also be confirmed from the Euler-Lagrange equations for the Euclidean action.)
In free space, bubbles are spheres in $D+1$ dimensions, with radii $R=D\sigma_{TF}/\epsilon$.

\subsection{Tunnelling exponents}

The bubble nucleation rate in a metastable state takes the exponential form given in Eqs. (\ref{Gamma1}) and (\ref{Gamma2})
with exponent $B$. Both vacuum and thermal tunnelling can be considered together if we introduce the speed of light
$c$ and make the following normalisations:
\begin{itemize}
\setlength{\itemsep}{0pt}
    \setlength{\parskip}{0pt}
\item $\sigma$ and $\epsilon$ are normalised by $\hbar c$ in the vacuum case
\item $\sigma$ and $\epsilon$ are normalised by $k_BT$ in the thermal case
\end{itemize}
In addition, the bubbles in the vacuum case have structure in the imaginary time direction.
We will retain the notation $D$ as the spatial dimension, so that the bubbles exist in $D+1$ dimensions in the vacuum case.

When walls are included, as in figure \ref{fig:walls}, the tunnelling exponent is given by the total energy
(or Euclidean action) difference between the system with a bubble and the system without a bubble,
\begin{equation}
B={\cal A}_W(\sigma_{ST}-\sigma_{SF})+{\cal A}_B\sigma_{FT}-\epsilon{\cal V}.
\end{equation}
Young's equation (\ref{young}) can be used to simplify this expression,
\begin{equation}
B=\left({\cal A}_B-{\cal A}_W\cos\theta\right)\sigma_{FT}-\epsilon{\cal V}\label{B}.
\end{equation}
Note that the contact angle $\theta$ is a fixed quantity and the radii of curvature will be extrema of the tunnelling exponent.

For a free spherical bubble not attached to a wall, ${\cal A}_W=0$ and the radii of curvature
are equal. The bubble radius $R$ and  tunnelling exponent $B_{b}$ for free bubbles in $D$ spatial dimensions are given by
\begin{align}
&\hbox{thermal case}&&R=\displaystyle\frac{(D-1)\sigma_{TF}}{\epsilon},&&
B_b=\displaystyle\frac{\epsilon{\cal V}_DR^{D}}{D-1}\label{rcritthermal},\\
&\hbox{vacuum case}&&R=\displaystyle\frac{D\sigma_{TF}}{\epsilon},&&
B_b=\displaystyle\frac{\epsilon{\cal V}_{D+1}R^{(D+1)}}{D}.\label{rcritvacuum}
\end{align}
where ${\cal V}_N$ is the volume of a unit sphere in $N$ dimensions.

\section{Spherical seeds}

The two possibilities are bubbles that surround the seed (interstitial) and bubbles that nucleate on the edge (edge).
The energy of spherical liquid droplets around seeds has been investigated previously in more down to earth contexts,
e.g \cite{doi:10.1021/la104628q}. 

\subsection{Interstitial case-thermal}

\begin{center}
\begin{figure}[ht]
\begin{center}\hfil
\scalebox{0.4}{\includegraphics{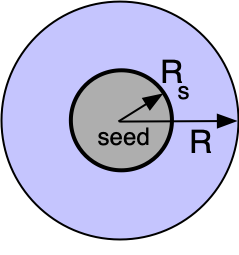}}\hfil
\scalebox{0.4}{\includegraphics{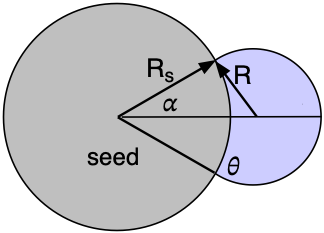}}
\end{center}
\caption{A bubble with interstitial seed (left) and a bubble on the edge of the seed (right).}
\label{fig:inter}
\end{figure}
\end{center}

\begin{center}
\begin{figure}[ht]
\begin{center}
\scalebox{0.25}{\includegraphics{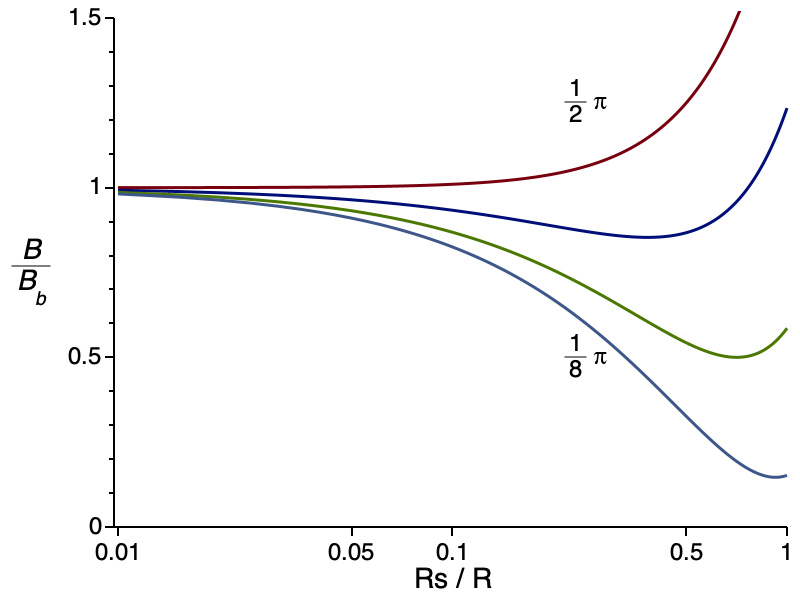}}
\scalebox{0.25}{\includegraphics{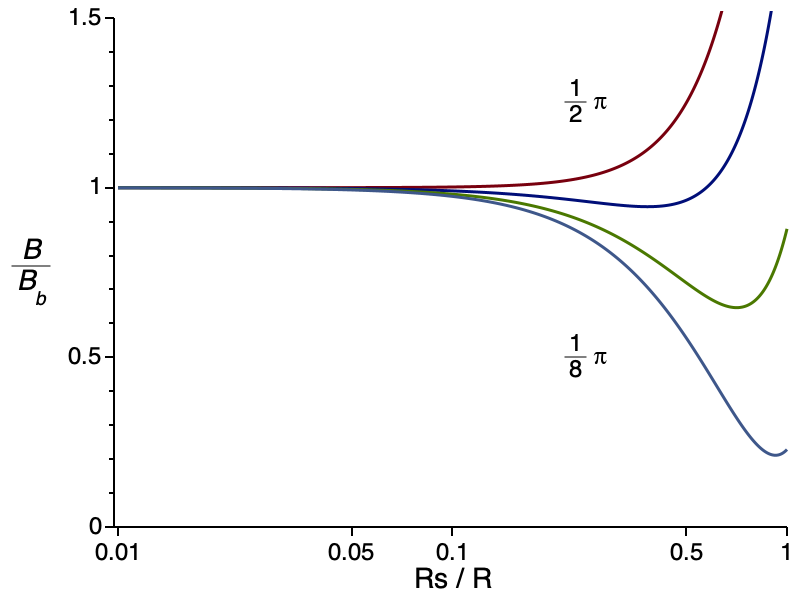}}
\end{center}
\caption{The thermal tunnelling exponent $B$ for interstitial seeds 
normalised by the free bubble case, as a function of the seed radius $R_s$
divided by the bubble radius $R=(D-1)\sigma_{TF}/\epsilon$. 
The curves are labeled by the angle of contact $\theta$, which is fixed by Young's relation
$\cos\theta=(\sigma_{SF}-\sigma_{ST})/\sigma_{TF}$.
The left-hand plot is in two spatial 
dimensions, $D=2$, and the right hand plot in three spatial dimensions, $D=3$.}
\label{fig:compare}
\end{figure}
\end{center}

Consider a bubble of radius $R$ and a seed of radius $R_s$. In the interstitial case, the bubble has
spherical symmetry with $R=(D-1)\sigma_{TV}/\epsilon$ from Eq. (\ref{rcritthermal}), and we have
\begin{align}
{\cal V}&={\cal V}_D(R^D-R_s^D),\\
{\cal A}_B&=D{\cal V}_DR^{D-1},\\
{\cal A}_W&=D{\cal V}_DR_s^{D-1},
\end{align}
where ${\cal V}_D$ is the volume inside a unit sphere in $D$ dimensions.
The exponent from Eq. (\ref{B}) is
\begin{equation}
B=B_{b}\left\{
1-\left(\frac{R_s}{R}\right)^{D-1}D\cos\theta+(D-1)\left(\frac{R_s}{R}\right)^D,
\right\},
\end{equation}
where the free tunnelling exponent $B_b$ is given in Eq.(\ref{rcritthermal}), and
the contact angle is fixed by Young's relation $\cos\theta=(\sigma_{SF}-\sigma_{ST})/\sigma_{TF}$.
The nucleation rate has a minimum with a seed radius $R_s=R\cos\theta$ when $\cos\theta>0$,
giving an enhanced nucleation rate, as shown in figure \ref{fig:compare}. When $\theta\ge \pi/2$, $B\ge B_{b}$ for any seed radius, and 
nucleation around the seed is always disfavoured.

\subsection{Edge case-thermal}

In the edge case, rotational symmetry about the axis joining the centre of the seed to the centre of the
bubble implies that the radii of curvature of the bubble have the same value $R$, and the bubble is part
of a sphere. The tunnelling exponent is given by the general formula Eq. (\ref{B}), where the areas and volumes 
can all be obtained from elementary Euclidean geometry. 

We introduce the angle $\alpha$ subtended by the bubble at the seed centre,
\begin{equation}
\tan\alpha=\frac{R\sin\theta}{R_s-R\cos\theta}.
\end{equation}
In two spatial dimensions,
\begin{align}
{\cal A}_W&=2R_s\alpha,\\
{\cal A}_B&=2R(\alpha+\theta),\\
{\cal V}&=R^2\left[(\alpha+\theta)-\frac12\sin2(\alpha+\theta)\right]
-R_s^2\left[\alpha-\frac12\sin 2\alpha\right].
\end{align}
The tunnelling exponent is found after eliminating $\alpha$. 
For example, with a contact angle $\theta=\pi/2$, the tunnelling exponent becomes
\begin{equation}
B=B_{b}\left\{ \left(\frac{1}{2}+\frac{1}{\pi}\arctan\frac{R}{R_s}\right)+\frac{1}{\pi}\frac{R_s^2}{R^2}\arctan\frac{R}{R_s}-
\frac{1}{\pi}\frac{R_s}{R}\right\},\label{BEdge}
\end{equation}
where $B_{b}=\epsilon\pi R^2$. In this case, the exponent runs from $B_{b}$ at $R_s=0$ down to 
$B_{b}/2$ as $R_s\to\infty$. Seeded nucleation is therefore favoured at $\theta=\pi/2$, 
unlike in the interstitial case.

\begin{center}
\begin{figure}[ht]
\begin{center}
\scalebox{0.25}{\includegraphics{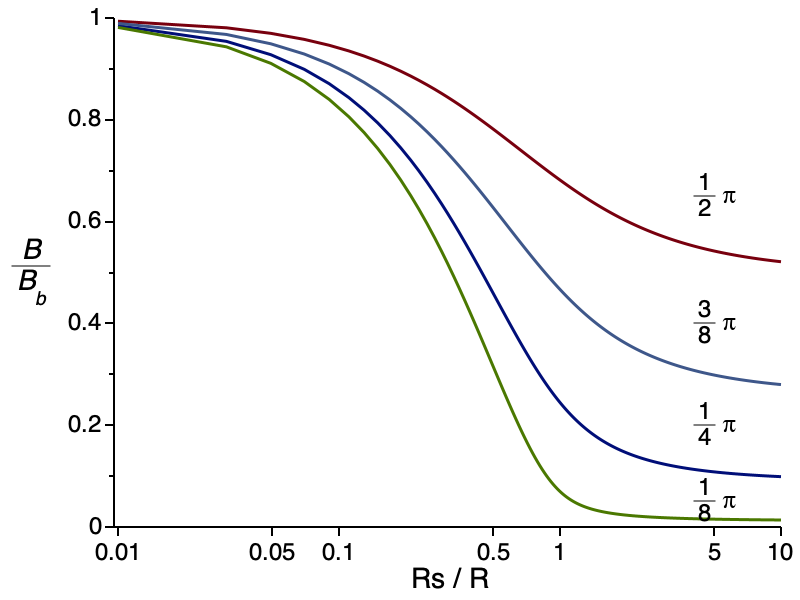}}
\scalebox{0.25}{\includegraphics{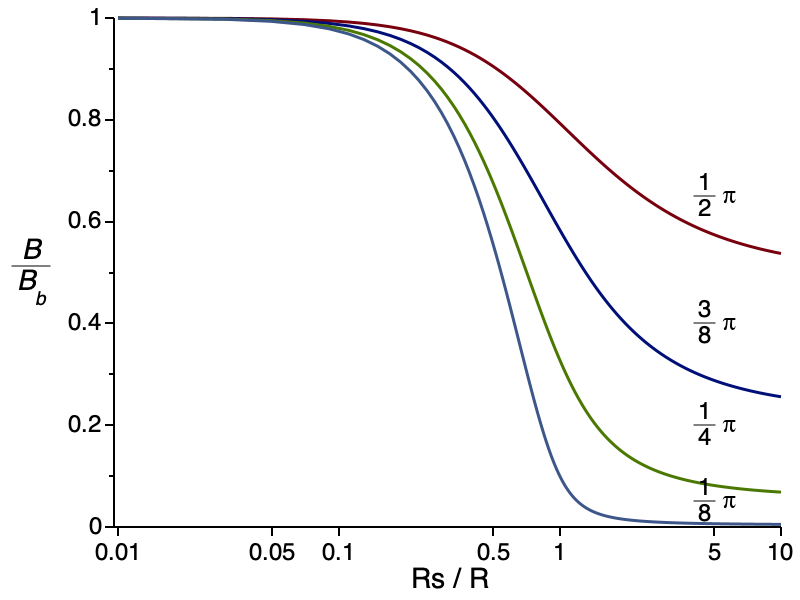}}
\end{center}
\caption{The thermal tunnelling exponent $B$ for the edge nucleation normalised by the free bubble case, as a function of the seed radius $R_s$ divided by the bubble radius $R=(D-1)\sigma_{TF}/\epsilon$. The curves are labeled by the angle of contact $\theta$, which is fixed by the surface tensions.
The left-hand plot is in two spatial dimensions, $D=2$, and the right hand plot in
three spatial dimensions, $D=3$.}
\label{fig:CompareEdge}
\end{figure}
\end{center}

For three spatial dimensions,
\begin{align}
{\cal A}_W&=2\pi R_s^2[1-\cos\alpha],\\
{\cal A}_B&=2\pi R^2[1-\cos(\alpha+\theta)],\\
{\cal V}&=\frac{\pi}{3}R^3\left[2-3\cos(\alpha+\theta)+\cos^3(\alpha+\theta)\right]
-\frac{\pi}{3}R_s^3\left[2-3\cos\alpha+\cos^3\alpha\right].
\end{align}
The tunnelling exponents in two and three spatial dimensions are show in figure \ref{fig:CompareEdge}, for a range of contact angles.
The tunnelling exponent is always smaller on the edge of a nucleation seed. The effect is most pronounced when the the seed
is larger than the bubble radius and the contact angle is small, and is likely to dominate over changes to the pre-factor discussed earlier. 
When the contact angle  $\theta=\pi/2$, and the seed radius is large, the tunnelling exponent is exactly half of the bulk case, 
corresponding to having half an instanton.

\subsection{Interstitial case-vacuum}

\begin{center}
\begin{figure}[ht]
\begin{center}
\scalebox{0.4}{\includegraphics{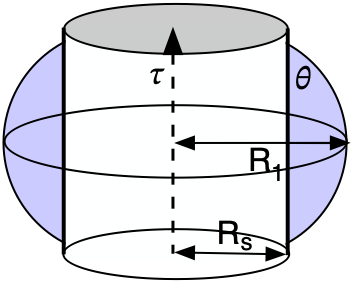}}
\scalebox{0.4}{\includegraphics{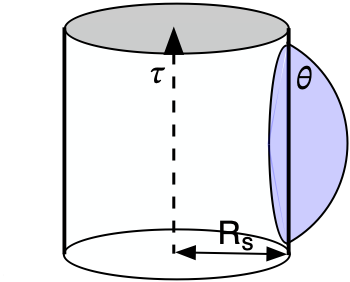}}
\end{center}
\caption{In the vacuum case, the seed extends into the imaginary time direction. The bubble wall meets the seed
with angle $\theta$, but the bubble radii of curvature are no longer equal. Left: interstitial case. Right: Edge case.}
\label{fig:vacuum}
\end{figure}
\end{center}

We restrict attention to Lorentz invariant systems with velocity of light $c$. 
The seed is a cylinder extended in the imaginary time direction. The
bubble wall meets the seed along the imaginary time direction, as shown in figure \ref{fig:vacuum}. 
The bubble wall is no longer spherical, but in the interstitial case it has cylindrical
symmetry with surface $r(\tau)$ in cylindrical coordinates. Imaginary time ranges from
$-\tau_s\le\tau\le\tau_s$, where $r(\tau_s)=R_s$.

\begin{center}
\begin{figure}[ht]
\begin{center}
\scalebox{0.25}{\includegraphics{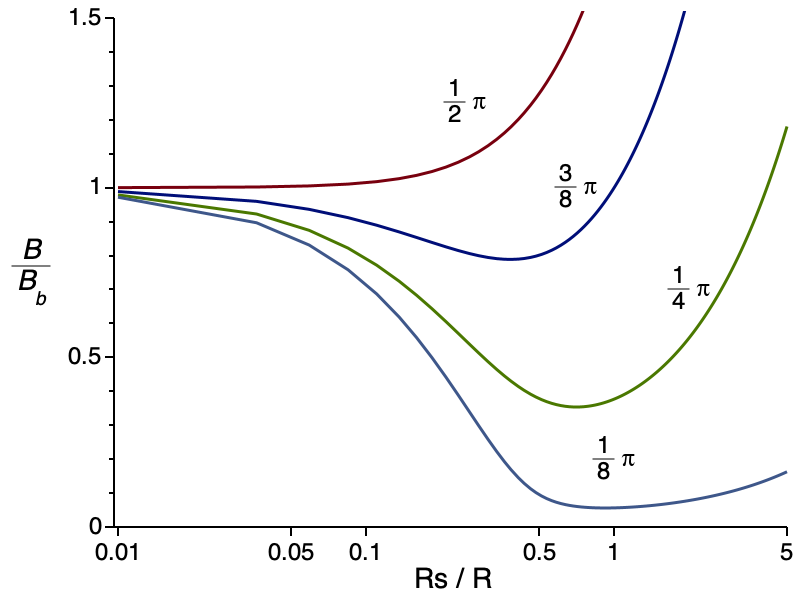}}
\scalebox{0.25}{\includegraphics{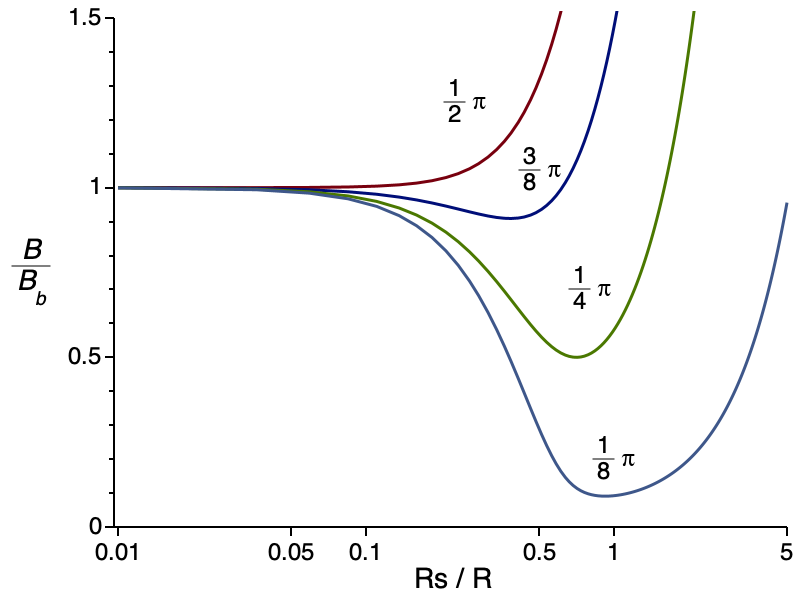}}
\end{center}
\caption{The vacuum tunnelling exponent $B$ for an interstitial seed  normalised by the free bubble case,  as a function of the seed radius $R_s$
normalised by the mean curvature radius $R=D\sigma_{TF}/\epsilon$. 
The curves are labeled by the angle of contact $\theta$, which is fixed by the surface tensions.
The left-hand plot is in two spatial dimensions, $D=2$, and the right hand plot in
three spatial dimensions, $D=3$.}
\label{fig:vaccompare}
\end{figure}
\end{center}

The bubble surface is a stationary point of the exponent $B$ given by Eq. (\ref{B}),
where the areas and volumes are given by
\begin{align}
{\cal A}_W&=2D{\cal V}_DR_s^{D-1}\int_0^{\tau_s} c\,d\tau,\\
{\cal A}_B&=2D{\cal V}_D\int_{0}^{\tau_s}c\,d\tau \,r^{D-1}(1+r^{\prime\,2}/c^2),\\
{\cal V}&=2{\cal V}_D\int_{0}^{\tau_s}c\,d\tau \,(r^D-R_s^D).
\end{align}
The Euler-Lagrange equations for the bubble surface have a first integral,
\begin{equation}
E=r^D-\frac{r^{D-1}}{(1+r^{\prime\,2}/c^2)^{1/2}}R.\label{first}
\end{equation}
The boundary condition $r'=c\tan\theta$ at the seed and $r'=0$ at the plane of symmetry $\tau=0$ imply
\begin{equation}
E=R_s^D-RR_s^{D-1}\cos\theta=R_1^D-RR_1^{D-1}.\label{boundary}
\end{equation}
Substituting the equation for the bubble wall (\ref{first}) into $B$ gives
\begin{equation}
B=2{\cal V}_D\epsilon\int_{R_s}^{R_1}dr\,\left[r^{2D-2}R^2-(E-r^D)^2\right]^{1/2},
\end{equation}
where $E$ and $R_1$ are determined by Eq. (\ref{boundary}) as functions of $R_s$ and $\theta$.
The result of evaluating the integral is shown for two and three spatial dimensions in figure \ref{fig:vaccompare}.
The tunnelling rate is considerably enhanced for small contact angles.
Unlike the thermal case, in the vacuum bubble case the seed can be larger than the mean curvature radius $R$.

\subsection{Edge case-vacuum}

For vacuum tunnelling in the edge case, the bubble sits on the edge of a cylinder, as in figure \ref{fig:vacuum}. 
The energy of bubbles with this geometry  has been investigated in three spatial dimensions \cite{doi:10.1021/la403088r}. 
Unfortunately, these results were 
presented for fixed volume instead of fixed ``pressure". Finding the bubble shape in general is a complex 
numerical problem  which we will  not pursue further here. However, we can obtain approximate results in the 
large seed radius limit, by adapting formulae from Ref. \cite{doi:10.1021/la403088r} (see appendix \ref{cylinders}).

In two spatial dimensions,
\begin{equation}
\frac{B}{B_b}=\frac12\left(1-\cos\theta-\frac12\cos\theta\sin^2\theta\right)
+\frac3{16}\frac{R}{R_s}\sin^4\theta
+O(R^2/R_s^2).
\end{equation}
In three spatial dimensions,
\begin{equation}
\frac{B}{B_b}= \frac{1}{\pi}\left(\theta-\sin\theta\cos\theta-\frac23\cos\theta\sin^3\theta\right)
+\frac{4}{5\pi}\frac{R}{R_s}\sin^5\theta+O(R^2/R_s^2).
\end{equation}
A useful check is that the instanton should become hemi-spherical in the limit $\theta=\pi/2$
and $R_s\to\infty$. Bubbles in the bulk are spherical, and we expect $B/B_b=1/2$, as is the case
from the formulae.

\begin{center}
\begin{figure}[ht]
\begin{center}
\scalebox{0.25}{\includegraphics{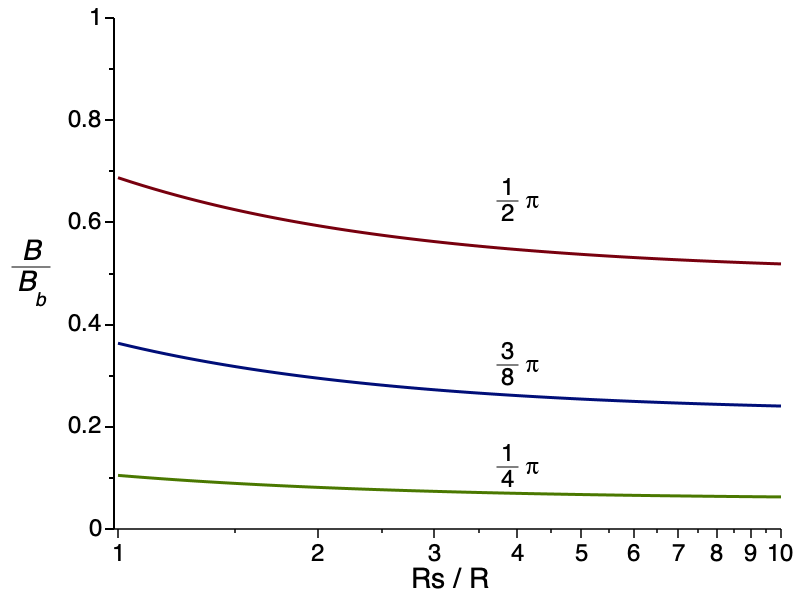}}
\scalebox{0.25}{\includegraphics{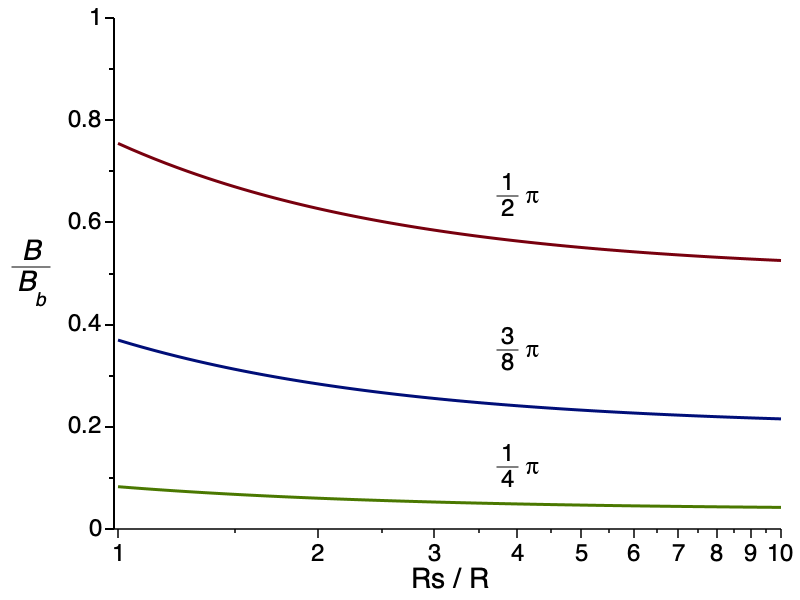}}
\end{center}
\caption{The vacuum tunnelling exponent $B$ for edge nucleation normalised by the free bubble case, 
as a function of the seed radius $R_s$ normalised by the mean curvature radius $R=D\sigma_{TF}/\epsilon$.
The curves are labelled by the contact angle $\theta$.
The left-hand plot is in two spatial dimensions, $D=2$ and the right hand plot in
three spatial dimensions, $D=3$.
}
\label{fig:edgevaccompare}
\end{figure}
\end{center}

The vacuum tunnelling exponents are plotted in figure \ref{fig:edgevaccompare}.
Compared to the interstitial case in figure \ref{fig:vaccompare}, these are much smaller.
Therefore, at least in this regime, vacuum bubble nucleation is predominantly seeded at the edge.

\section{Analogue systems}

The analogue system we consider here is similar to the one described in Ref \cite{Jenkins:2023npg}.
It is based on potassium atoms occupying two hyperfine levels condensed in a two-dimensional atom trap.
We regard the atoms in the two levels as two separate components of the BEC.
Atomic collisions between atoms in the same level, or different levels, are described by three parameters
$g_{11}$, $g_{22}$ and $g_{12}$. 
In addition, a modulated  microwave field provides mixing between the atoms in each level, 
described by a Rabi frequency
$\Omega$ and a dimensionless parameter $\lambda$.
The scattering parameters $g_{ij}$ determine the relative number density of the two components in the
ground state of the system, $n_1$ and $n_2$. 
Important physical parameters are the frequency scale $\omega_m=(g_{11}+g_{22}-2g_{12})n/\hbar$
and healing length $\xi_m=(\hbar/m\omega_m)^{1/2}$, where $n=n_1+n_2$ is the total density.

An additional trapping potential $V_T(x,y)$  is used as a nucleation seed. Inside the seed, the potential drives the
density to zero. There is a narrow transition region at the edge of the seed, which we can arrange to 
have a width approximately equal to $\xi_m$. The density outside the seed has a constant value $n$
in the initial state.

In  \cite{Jenkins:2023npg}, it was shown that the system can be described by an effective theory for a single
scalar field related to the relative phase $\varphi$ of the two components.
The canonically normalised field $\phi=v \varphi$, where
\begin{equation}
v^2=n_1n_2\xi_m^2\hbar\omega_m /n.
\end{equation}
The field equation has Klein-Gordon form,
\begin{equation}
c_\varphi^{-2}\ddot\phi-\nabla^2\phi+\frac{dV}{d\phi}=0,
\end{equation}
where the sound speed $c_\varphi=\xi_m\omega_m(n_1n_2)^{1/2}\sqrt{2}/n$ and the potential
\begin{equation}
V(\phi)=\hbar\Omega \sqrt{n_1n_2}\left(-\cos\frac\phi{v}+\frac12(\lambda^2-1)\sin^2\frac\phi{v}\right).\label{pot}
\end{equation}
When $\lambda^2>1$, the potential has a local minimum, or false vacuum, at $\phi=v\pi$ and a global minimum, or true vacuum,
at $\phi=0$. The energy density difference between the vaccua is $\epsilon=2\hbar\Omega\sqrt{n_1n_2}$.

For the finite temperature simulations, we solve the Projected Stochastic Gross-Pitaevski Equation (SPGPE) for
the condensate fields $\psi_i$ \cite{BradleyStochastic2014}. The SPGPE is
\begin{equation}
i\hbar\frac{\partial\psi_i}{\partial t}={\cal P}\left\{(1+i\gamma)\frac{\partial H}{\partial \overline\psi_i}+\eta_i\right\},
\end{equation}
where $H$ is the Hamiltonian, $i\gamma$ is a dissipation term and $\eta_m$ is a Gaussian stochastic noise term
with statistics
\begin{equation}
\langle\eta_i({\bf r},t)\eta_j^\dagger({\bf r}',t')\rangle=\frac{2\gamma k_BT}{\hbar\omega_m n}\delta_{ij}\delta({\bf r}-{\bf r}')\delta(t-t').
\end{equation}
The equation is solved in a two dimensional periodic box with the trapping potential seed at the centre. Averages
over many runs are used to find the bubble nucleation rate for a range of seed sizes.

The system is initialised in the false vacuum state. This is achieved by equilibrating in the true vacuum state
and changing the sign of the Rabi frequency $\Omega$ with a piecewise linear ramp. This switches the
true and false vacua. Bubble nucleation times are evaluated relative to the end of the ramp. The
nucleation rate is extracted by fitting the nucleation times to a Poisson distribution. (A small time offset
parameter is included to allow for the bubble nucleation detection algorithm and nucleation during the ramp.)

\begin{center}
\begin{figure}[ht]
\begin{center}
\scalebox{0.27}{\includegraphics{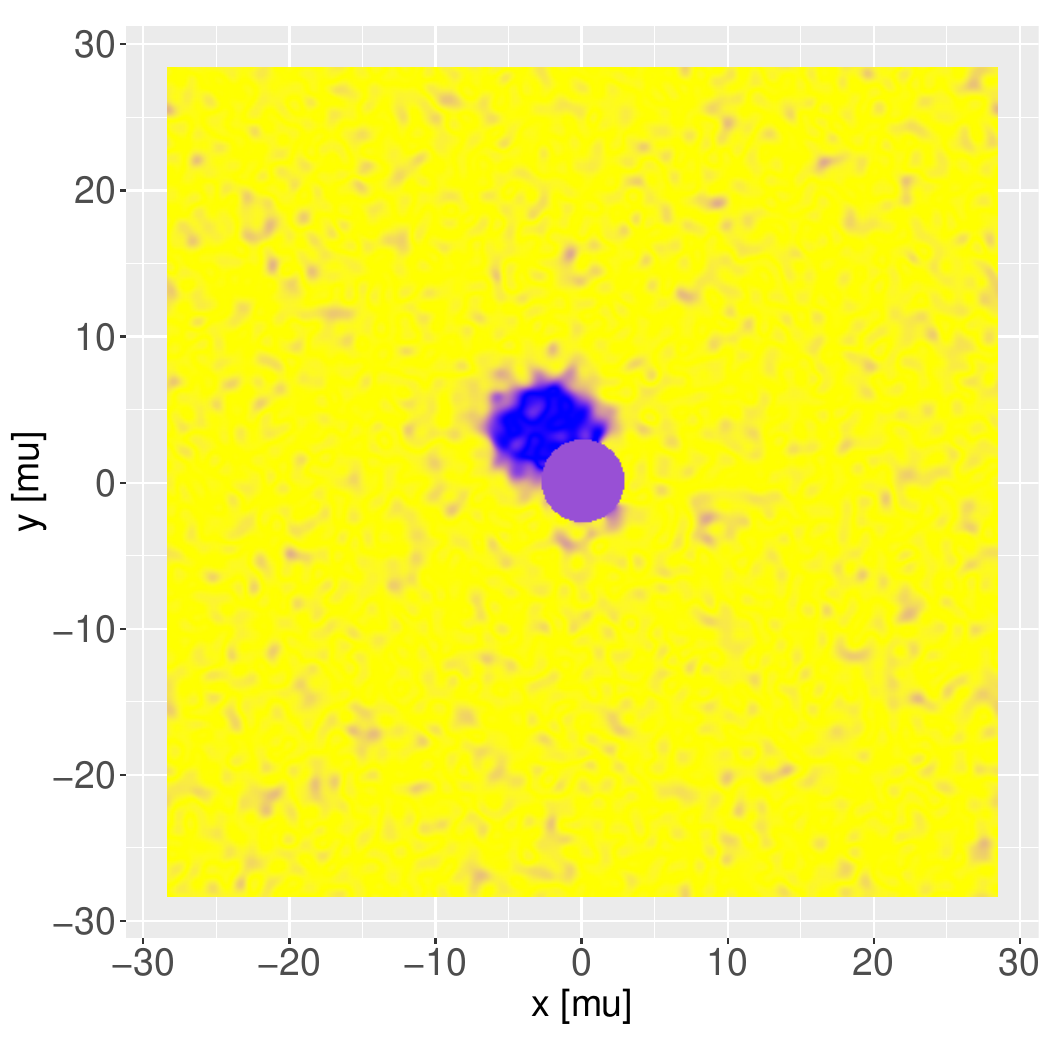}}
\scalebox{0.27}{\includegraphics{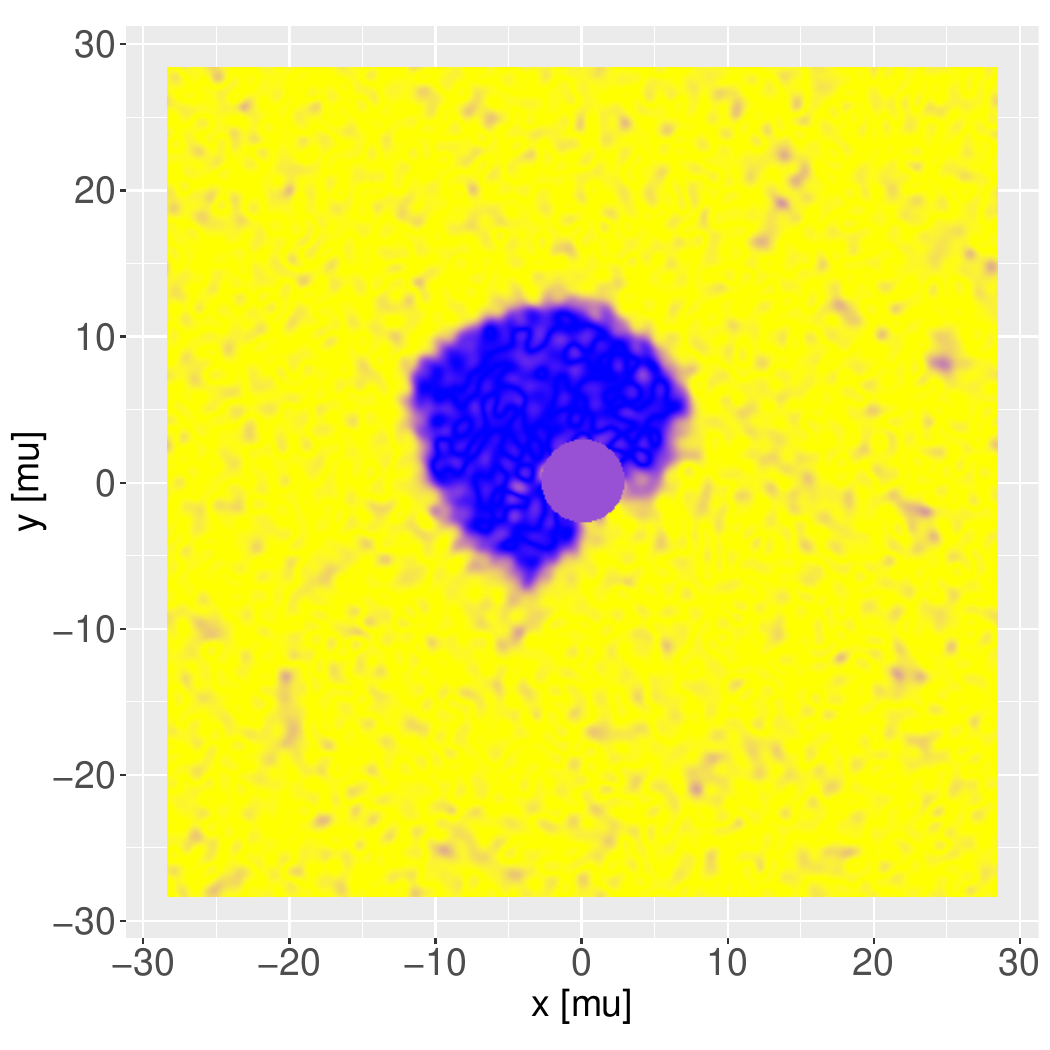}}
\scalebox{0.27}{\includegraphics{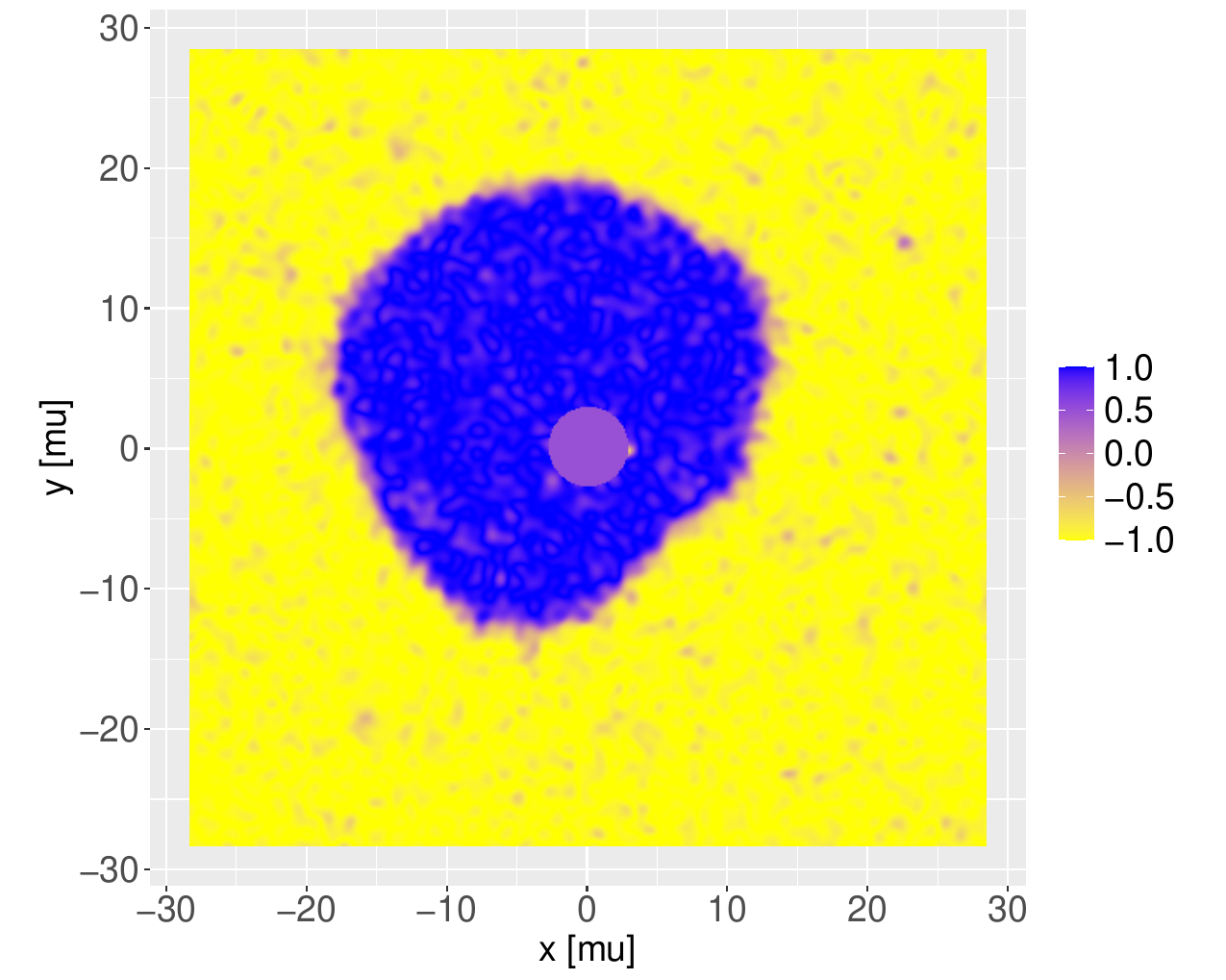}}
\end{center}
\caption{Simulated thermal bubble nucleation around a circular seed for a two dimensional BEC. The plot shows
the cosine of the relative phase of the two components. The seed is the grey circle in the centre. 
Snapshots are taken at $28{\rm ms}$, $29{\rm ms}$ and $30{\rm ms}$. This 39K model has $200$ 
atoms per square micron, temperature
$160{\rm nK}$, healing length $\xi_m=0.142\,\mu{\rm m}$, frequency $\omega_m=14.0\times2\pi\,{\rm kHz}$ and 
Rabi frequency $\Omega=100\times2\pi\,{\rm Hz}$.
}
\label{fig:bubblepic}
\end{figure}
\end{center}

Bubble nucleation in a typical run is shown in figure \ref{fig:bubblepic}. The pictures show the cosine
of $\phi/v$, where $\phi$ is the degree of freedom of the effective field theory. In the false vacuum phase
$\cos(\phi/v)=-1$, and in the true vacuum phase $\cos(\phi/v)=1$. The field $\phi$ is not defined
inside the seed where the atomic density vanishes. Bubbles of true vacuum phase
nucleate, as expected, on the edge of the seed, with contact angle is around $\pi/2$. They grow at the 
sound speed for the BEC.

For comparison with the theory presented earlier, we fit the seeded nucleation rate $\Gamma_s$ to the formula
\begin{equation}
\Gamma_s=A_s R_s B_s^{1/2}e^{-B_s},
\end{equation}
where $B_s\equiv B_s(B_b,R,R_s)$ is given by Eq. (\ref{BEdge}). For small seeds, the bulk nucleation
rate $\Gamma_b$ can exceed the seeded rate. The bulk nucleation rate depends on the size $L$ of the periodic box,
\begin{equation}
\Gamma_b=A_b L^2 B_be^{-B_b}.
\end{equation}
The total nucleation rate $\Gamma=\Gamma_s+\Gamma_b$ can be fit with four parameters 
$A_s$, $A_b$, $B_b$ and $R$. (In practice, we reduced to three parameters by fixing the $R_s=0$
point).  The rates are shown in  figure \ref{fig:ratefit}, assuming the contact angle of $\theta=\pi/2$
as predicted by theory (see appendix \ref {apB}), and seen in pictures such as figure \ref{fig:bubblepic}. The results suggest that the theory is a good
description of seeded nucleation, even though the theory uses a thin-wall approximation and the 
actual bubble walls are quite broad.

\begin{center}
\begin{figure}[ht]
\begin{center}
\scalebox{0.5}{\includegraphics{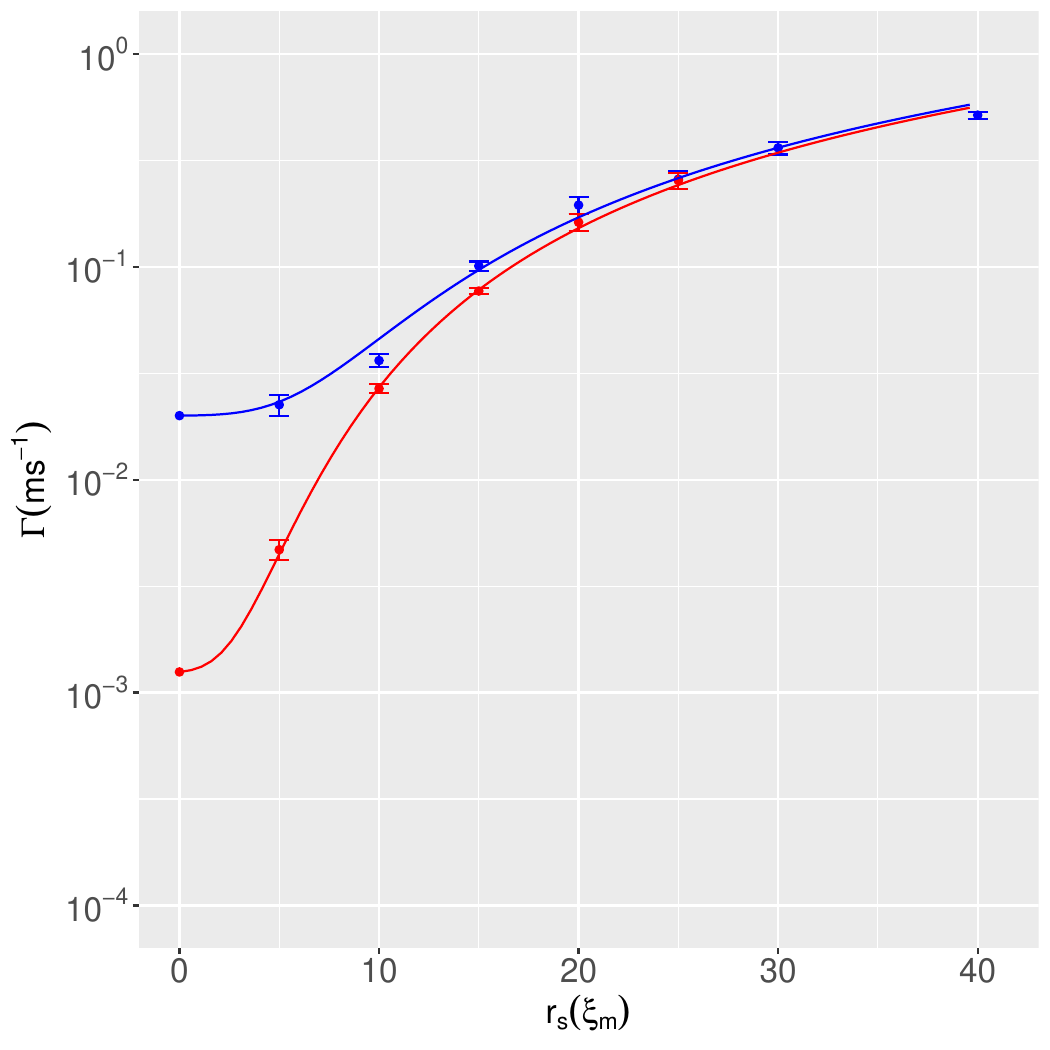}}
\end{center}
\caption{Bubble nucleation rate $\Gamma$ plotted as a function of the seed radius $R_s$ in healing
length units. The upper curve is for a periodic box of side $L=400\xi_m$ and the lower curve
$L=200\xi_m$. The curves are a fit to the data using the theory described in the text. The fitting
parameters are bubble radius
 $R=9.2\xi_m$ and free nucleation exponent $B_b=14.7$. The contact angle $\theta=\pi/2$.
}
\label{fig:ratefit}
\end{figure}
\end{center}

\section{conclusion}

We have presented fairly comprehensive results for tunnelling exponents in false vacuum decay
at zero and nonzero temperature around a spherical nucleation seed. We have looked at both two
and three spatial dimensions. Whilst the three dimensional case is the one applicable to the universe,
two dimensions have applications to analogue false vacuum decay experiments. In this context, we
have compared the thermal nucleation rates to a numerical simulation of the time evolution of a
Bose Einstein condensate.
There is good agreement between the instanton and real-time approaches, even though the
bubbles in this system don't have the thin walls used in the instanton theory. It would be fascinating
to compare the instanton theory to the vacuum nucleation in a real experiment.

The form that nucleation seeds might take in an early universe application is an open question.
Our results do not apply directly to some of the proposals on nucleation that have been put forward. 
However, it is tempting to speculate. Black hole seeding of bubble nucleation has only
been looked at with spherical symmetry so far \cite{PhysRevD.32.1333,Gregory_2014}. 
Our results have a definite preference for bubbles
appearing on the edge of seed, which is not a spherically symmetric configuration.

In the case of proton or heavy ion collisions, the initial collision region is likeley to be pancake shaped due to
the Lorentz contraction of the particles. In the right conditions, the energy can redistribute itself
into a more spherical region, with fluid-like properties. This is not so dissimilar to the seeds we have considered
here, and it may be possible to define surface energy or actions to apply the general theory.

\section*{Data availability statement}

Data supporting this publication are openly available under a Creative Commons CC-BY-4.0 License in \cite{data}

\acknowledgments
The authors are grateful for discussion with Tom Billam, Kate Brown and Alex Jenkins.
This work is supported by the UK Science and Technology Facilities Council
[grants ST/T00584X/1 and ST/W006162/1].

\appendix

\section{Bubbles on cylinders}\label{cylinders}

This appendix gives the mapping of functions from the liquid droplet problem in Ref. \cite{doi:10.1021/la403088r}
to the vacuum bubble nucleating on the edge of a seed in two spatial dimensions. The relevant functions 
defined in Ref. \cite{doi:10.1021/la403088r} are
\begin{align}
A_0(r)&=\frac14 r^2,\\
C_0(\theta)&=-\frac14\sin^2\theta.
\end{align}
In addition, the droplet problem parameterises changes in $R$ by a constant $k$. We have fixed $R$, so that $k=0$ in
our application. Quoting Eqs (24), (38) and (39) 
\footnote{In Eq. (39) of Ref. \cite{doi:10.1021/la403088r}, $\cos\alpha$ should read $\cos\alpha\sin\alpha$.}
from  \cite{doi:10.1021/la403088r}, 
and putting in the correct scaling by the mean curvature radius $R$,
\begin{align}
{\cal V}&=\frac\pi3R^3\left(1-\cos\theta-\frac12\cos\theta\sin^2\theta\right)
+\frac{3\pi}{8}R^4R_s^{-1}\sin^4\theta+O(R^5R_s^{-2}),\\
{\cal A}_B&=2\pi R^2(1-\cos\theta)+\pi R^3 R_s^{-1}\sin^2\theta+O(R^5R_s^{-2}),\\
{\cal A}_W&=\pi R^2\sin^2\theta+\pi R^3 R_s^{-1}\cos\theta\sin^2\theta++O(R^5R_s^{-2}).
\end{align}
These can be used in Eq. (\ref{B}) with $R=2\sigma_{TF}/\epsilon$ to eliminate $\sigma_{TF}$,
\begin{equation}
B=\frac\pi3\epsilon R^3\left(1-\cos\theta-\frac12\cos\theta\sin^2\theta\right)
+\frac\pi8\epsilon R^4R_s^{-1}\sin^4\theta+O(R^5R_s^{-2}).
\end{equation}
For comparison, the bulk tunnelling exponent from Eq. (\ref{rcritvacuum})  is $B_b=2\pi\epsilon R^3/3$.

Extending the result to three spatial dimensions simply involves replacing the area measure
$2\pi r dr$ by $4\pi r^2 dr$. The volumes and areas become
\begin{align}
{\cal V}&=\frac\pi2R^4\left(\theta-\sin\theta\cos\theta-\frac23\cos\theta\sin^3\theta\right)
+\frac{8\pi}{15}R^5R_s^{-1}\sin^5\theta+O(R^6R_s^{-2}),\\
{\cal A}_B&=2\pi R^3(\theta-\cos\theta\sin\theta)+2\pi R^4 R_s^{-1}\sin^3\theta+O(R^6R_s^{-2}),\\
{\cal A}_W&=\frac{4\pi}{3}R^3\sin^3\theta+2\pi R^4 R_s^{-1}\cos\theta\sin^3\theta+O(R^6R_s^{-2}).
\end{align}
Using Eq. (\ref{B}) with $R=3\sigma_{TF}/\epsilon$ gives
\begin{equation}
B=\frac\pi6\epsilon R^4\left(\theta-\sin\theta\cos\theta-\frac23\cos\theta\sin^3\theta\right)
+\frac{2\pi}{15}\epsilon R^5R_s^{-1}\sin^5\theta+O(R^6R_s^{-2}).
\end{equation}
The bulk tunnelling exponent is $B_b=\pi^2\epsilon R^4/6$.

\section{Contact angle for the analogue system}\label{apB}

The two state system in two dimensions with condensate
field $\psi=\{\psi_1,\psi_2\}$, and trapping potential $V_T$, has Hamiltonian
\begin{equation}
H=\int\left\{-\frac{\hbar^2}{2m}\psi^\dagger\nabla^2\psi
+\frac12\sum_{i,j} g_{ij}|\psi_i|^2|\psi_j|^2+
(V_T-\mu)\psi^\dagger \psi
-\frac{\hbar\Omega}{2}\psi^\dagger\sigma_x \psi
+\frac{\hbar\Omega}2g'(\psi^\dagger\sigma_y\psi)^2,
\right\}dxdy,\label{vstat}
\end{equation}
where $g'=\lambda^2/4\sqrt{n_1n_2}$, the $\sigma$ are Pauli matrices, 
and other parameters were defined in the main text.
We use a Bloch sphere representation for the fields,
\begin{align}
\psi_1&=\sqrt{n}\cos\frac{\theta}{2}e^{i(\phi+\varphi)/2},\\
\psi_2&=\sqrt{n}\sin\frac{\theta}{2}e^{i(\phi-\varphi)/2}.
\end{align}
In Ref. \cite{Jenkins:2023npg}, it was shown that the dynamics
of bubble nucleation can be recovered from an effective theory using the relative phase
$\varphi$ with the potential (\ref{pot}).

To find the surface tension at the wall we consider a plane wall along the $y$ axis, with number density $n\equiv n(x)$, and 
calculate the energy per unit length in the $y$ direction. (The equations for the angular variables imply that
they remain constant.) 
The Hamiltonian per unit length ${\cal H}$ becomes
\begin{equation}
{\cal H}=\int_0^\infty\left\{
\frac{\hbar^2}{2m}(\nabla\sqrt{n})^2+\frac12\hat g n^2+(V_T-\mu)n\pm\frac{\hbar}{2}\Omega n\sin\theta
\right\}dx,
\end{equation}
with upper and lower signs for the false and true vacuum respectively, and $\hat g=n\det g_{ij}/(4\hbar\omega_m)$. 
Variation of the Hamiltonian enables us to eliminate the chemical potential in favour of the constant vacuum density $\bar n$ at $V_T=0$,
\begin{equation}
{\cal H}=\int_0^\infty\left\{
\frac{\hbar^2}{2m}(\nabla\sqrt{n})^2+\frac12\hat g n(n-2\bar n)+V_Tn
\right\}dx.
\end{equation}
Note that the Rabi frequency dependence is now in $\bar n$, and there are two possible choices
$\bar n=\bar n_{TV}$ and $\bar n=\bar n_{FV}$ that differ by
$\bar n_{TV}-\bar n_{FV}=\hbar\Omega/\hat g$. 
For the surface tension at the wall, we have to subtract the
constant density term,
\begin{equation}
\sigma={\cal H}(n)-{\cal H}(\bar n)=
\int_0^\infty\left\{
\frac{\hbar^2}{2m}(\nabla\sqrt{n})^2+\frac12\hat g (n-\bar n)^2+V_T(n-\bar n)
\right\}dx.
\end{equation}
From this, we get the field equation for $n$,
\begin{equation}
\frac{\hbar^2}{2m}n^{-1/2}\nabla^2 n^{1/2}=\hat g(n-\bar n)+V_T.
\end{equation}
The simplest case for a step, $V_T=0$ for $x>0$, has $n=\bar n\tanh^2(x/\xi)$ where the healing length
$\xi=\hbar/(2m\hat g\bar n)^{1/2}$.
The integral gives
\begin{equation}
\sigma=\frac23\hat g\xi \bar n^2.
\end{equation}
However, the relevant quantity for calculating the contact angle $\theta$ is the difference $\sigma_{SF}-\sigma_{ST}$.
We replace $\bar n$ by the value in the respective vacua, $\bar n_{FV}$ or $\bar n_{TV}$,
\begin{equation}
\sigma_{SF}-\sigma_{ST}=\frac23\hat g\xi_{FV} \bar n_{FV}^2-\frac23\hat g\xi_{TV} \bar n_{TV}^2
\approx -\xi_{FV}\bar n_{FV}\hbar\Omega,
\end{equation}
where we have used $n_{TV}-n_{FV}=\hbar\Omega/\hat g$.

For the bubble wall, $\sigma_{TF}$, the thin wall approximation suggests a value determined by the integral of 
$[V(\phi)-V(\phi_{FV})]^{1/2}$, which
results in $\sigma_{TF}=O(\Omega/\omega_m)^{1/2}$. Alternatively, the numerical
tunnelling exponents for thick wall bubbles from Ref. \cite{Abed:2020lcf} imply $B_b=O(\Omega/\omega_m)$. 
If we match this to the thin wall result (\ref{rcritthermal}), we obtain $\sigma_{TF}=O(\Omega/\omega_m)^0$. 
In either case, we deduce from Young's equation (\ref{young}) that $\cos\theta\approx0$ for small 
$\Omega$, and $\theta\approx\pi/2$.

\bibliography{NuclSeeds.bib}

\begin{thebibliography}{39}%
\makeatletter
\providecommand \@ifxundefined [1]{%
 \@ifx{#1\undefined}
}%
\providecommand \@ifnum [1]{%
 \ifnum #1\expandafter \@firstoftwo
 \else \expandafter \@secondoftwo
 \fi
}%
\providecommand \@ifx [1]{%
 \ifx #1\expandafter \@firstoftwo
 \else \expandafter \@secondoftwo
 \fi
}%
\providecommand \natexlab [1]{#1}%
\providecommand \enquote  [1]{``#1''}%
\providecommand \bibnamefont  [1]{#1}%
\providecommand \bibfnamefont [1]{#1}%
\providecommand \citenamefont [1]{#1}%
\providecommand \href@noop [0]{\@secondoftwo}%
\providecommand \href [0]{\begingroup \@sanitize@url \@href}%
\providecommand \@href[1]{\@@startlink{#1}\@@href}%
\providecommand \@@href[1]{\endgroup#1\@@endlink}%
\providecommand \@sanitize@url [0]{\catcode `\\12\catcode `\$12\catcode
  `\&12\catcode `\#12\catcode `\^12\catcode `\_12\catcode `\%12\relax}%
\providecommand \@@startlink[1]{}%
\providecommand \@@endlink[0]{}%
\providecommand \url  [0]{\begingroup\@sanitize@url \@url }%
\providecommand \@url [1]{\endgroup\@href {#1}{\urlprefix }}%
\providecommand \urlprefix  [0]{URL }%
\providecommand \Eprint [0]{\href }%
\providecommand \doibase [0]{http://dx.doi.org/}%
\providecommand \selectlanguage [0]{\@gobble}%
\providecommand \bibinfo  [0]{\@secondoftwo}%
\providecommand \bibfield  [0]{\@secondoftwo}%
\providecommand \translation [1]{[#1]}%
\providecommand \BibitemOpen [0]{}%
\providecommand \bibitemStop [0]{}%
\providecommand \bibitemNoStop [0]{.\EOS\space}%
\providecommand \EOS [0]{\spacefactor3000\relax}%
\providecommand \BibitemShut  [1]{\csname bibitem#1\endcsname}%
\let\auto@bib@innerbib\@empty
\bibitem [{\citenamefont {Coleman}(1977)}]{Coleman:1977py}%
  \BibitemOpen
  \bibfield  {author} {\bibinfo {author} {\bibfnamefont {Sidney~R.}\
  \bibnamefont {Coleman}},\ }\bibfield  {title} {\enquote {\bibinfo {title}
  {{The Fate of the False Vacuum. 1. Semiclassical Theory}},}\ }\href {\doibase
  10.1103/PhysRevD.15.2929, 10.1103/PhysRevD.16.1248} {\bibfield  {journal}
  {\bibinfo  {journal} {Phys. Rev.}\ }\textbf {\bibinfo {volume} {D15}},\
  \bibinfo {pages} {2929--2936} (\bibinfo {year} {1977})},\ \bibinfo {note}
  {[Erratum: Phys. Rev.D16,1248(1977)]}\BibitemShut {NoStop}%
\bibitem [{\citenamefont {Callan}\ and\ \citenamefont
  {Coleman}(1977)}]{Callan:1977pt}%
  \BibitemOpen
  \bibfield  {author} {\bibinfo {author} {\bibfnamefont {Curtis~G.}\
  \bibnamefont {Callan}}\ and\ \bibinfo {author} {\bibfnamefont {Sidney~R.}\
  \bibnamefont {Coleman}},\ }\bibfield  {title} {\enquote {\bibinfo {title}
  {{The Fate of the False Vacuum. 2. First Quantum Corrections}},}\ }\href
  {\doibase 10.1103/PhysRevD.16.1762} {\bibfield  {journal} {\bibinfo
  {journal} {Phys. Rev.}\ }\textbf {\bibinfo {volume} {D16}},\ \bibinfo {pages}
  {1762--1768} (\bibinfo {year} {1977})}\BibitemShut {NoStop}%
\bibitem [{\citenamefont {Linde}(1983)}]{1983544}%
  \BibitemOpen
  \bibfield  {author} {\bibinfo {author} {\bibfnamefont {A.}~\bibnamefont
  {Linde}},\ }\bibfield  {title} {\enquote {\bibinfo {title} {Decay of the
  false vacuum at finite temperature},}\ }\href {\doibase
  https://doi.org/10.1016/0550-3213(83)90072-X} {\bibfield  {journal} {\bibinfo
   {journal} {Nuclear Physics B}\ }\textbf {\bibinfo {volume} {223}},\ \bibinfo
  {pages} {544} (\bibinfo {year} {1983})}\BibitemShut {NoStop}%
\bibitem [{\citenamefont {Shaposhnikov}(1987)}]{SHAPOSHNIKOV1987757}%
  \BibitemOpen
  \bibfield  {author} {\bibinfo {author} {\bibfnamefont {M.E.}\ \bibnamefont
  {Shaposhnikov}},\ }\bibfield  {title} {\enquote {\bibinfo {title} {Baryon
  asymmetry of the universe in standard electroweak theory},}\ }\href {\doibase
  https://doi.org/10.1016/0550-3213(87)90127-1} {\bibfield  {journal} {\bibinfo
   {journal} {Nuclear Physics B}\ }\textbf {\bibinfo {volume} {287}},\ \bibinfo
  {pages} {757--775} (\bibinfo {year} {1987})}\BibitemShut {NoStop}%
\bibitem [{\citenamefont {Hogan}(1986)}]{10.1093/mnras/218.4.629}%
  \BibitemOpen
  \bibfield  {author} {\bibinfo {author} {\bibfnamefont {C.~J.}\ \bibnamefont
  {Hogan}},\ }\bibfield  {title} {\enquote {\bibinfo {title} {{Gravitational
  radiation from cosmological phase transitions}},}\ }\href {\doibase
  10.1093/mnras/218.4.629} {\bibfield  {journal} {\bibinfo  {journal} {Monthly
  Notices of the Royal Astronomical Society}\ }\textbf {\bibinfo {volume}
  {218}},\ \bibinfo {pages} {629--636} (\bibinfo {year} {1986})},\ \Eprint
  {http://arxiv.org/abs/https://academic.oup.com/mnras/article-pdf/218/4/629/3299141/mnras218-0629.pdf}
  {https://academic.oup.com/mnras/article-pdf/218/4/629/3299141/mnras218-0629.pdf}
  \BibitemShut {NoStop}%
\bibitem [{\citenamefont {Vilenkin}(1982)}]{Vilenkin:1982de}%
  \BibitemOpen
  \bibfield  {author} {\bibinfo {author} {\bibfnamefont {Alexander}\
  \bibnamefont {Vilenkin}},\ }\bibfield  {title} {\enquote {\bibinfo {title}
  {{Creation of Universes from Nothing}},}\ }\href {\doibase
  10.1016/0370-2693(82)90866-8} {\bibfield  {journal} {\bibinfo  {journal}
  {Phys. Lett. B}\ }\textbf {\bibinfo {volume} {117}},\ \bibinfo {pages}
  {25--28} (\bibinfo {year} {1982})}\BibitemShut {NoStop}%
\bibitem [{\citenamefont {Coleman}\ and\ \citenamefont
  {De~Luccia}(1980)}]{Coleman:1980aw}%
  \BibitemOpen
  \bibfield  {author} {\bibinfo {author} {\bibfnamefont {Sidney~R.}\
  \bibnamefont {Coleman}}\ and\ \bibinfo {author} {\bibfnamefont {Frank}\
  \bibnamefont {De~Luccia}},\ }\bibfield  {title} {\enquote {\bibinfo {title}
  {{Gravitational Effects on and of Vacuum Decay}},}\ }\href {\doibase
  10.1103/PhysRevD.21.3305} {\bibfield  {journal} {\bibinfo  {journal} {Phys.
  Rev. D}\ }\textbf {\bibinfo {volume} {21}},\ \bibinfo {pages} {3305}
  (\bibinfo {year} {1980})}\BibitemShut {NoStop}%
\bibitem [{\citenamefont {Mazumdar}\ and\ \citenamefont
  {White}(2019)}]{Mazumdar:2018dfl}%
  \BibitemOpen
  \bibfield  {author} {\bibinfo {author} {\bibfnamefont {Anupam}\ \bibnamefont
  {Mazumdar}}\ and\ \bibinfo {author} {\bibfnamefont {Graham}\ \bibnamefont
  {White}},\ }\bibfield  {title} {\enquote {\bibinfo {title} {{Review of cosmic
  phase transitions: their significance and experimental signatures}},}\ }\href
  {\doibase 10.1088/1361-6633/ab1f55} {\bibfield  {journal} {\bibinfo
  {journal} {Rept. Prog. Phys.}\ }\textbf {\bibinfo {volume} {82}},\ \bibinfo
  {pages} {076901} (\bibinfo {year} {2019})},\ \Eprint
  {http://arxiv.org/abs/1811.01948} {arXiv:1811.01948 [hep-ph]} \BibitemShut
  {NoStop}%
\bibitem [{\citenamefont {Hindmarsh}\ \emph {et~al.}(2021)\citenamefont
  {Hindmarsh}, \citenamefont {Lüben}, \citenamefont {Lumma},\ and\
  \citenamefont {Pauly}}]{10.21468/SciPostPhysLectNotes.24}%
  \BibitemOpen
  \bibfield  {author} {\bibinfo {author} {\bibfnamefont {Mark}\ \bibnamefont
  {Hindmarsh}}, \bibinfo {author} {\bibfnamefont {Marvin}\ \bibnamefont
  {Lüben}}, \bibinfo {author} {\bibfnamefont {Johannes}\ \bibnamefont
  {Lumma}}, \ and\ \bibinfo {author} {\bibfnamefont {Martin}\ \bibnamefont
  {Pauly}},\ }\bibfield  {title} {\enquote {\bibinfo {title} {{Phase
  transitions in the early universe}},}\ }\href {\doibase
  10.21468/SciPostPhysLectNotes.24} {\bibfield  {journal} {\bibinfo  {journal}
  {SciPost Phys. Lect. Notes}\ ,\ \bibinfo {pages} {24}} (\bibinfo {year}
  {2021})}\BibitemShut {NoStop}%
\bibitem [{\citenamefont {Ghan}\ \emph {et~al.}(2011)\citenamefont {Ghan},
  \citenamefont {Abdul-Razzak}, \citenamefont {Nenes}, \citenamefont {Ming},
  \citenamefont {Liu}, \citenamefont {Ovchinnikov}, \citenamefont {Shipway},
  \citenamefont {Meskhidze}, \citenamefont {Xu},\ and\ \citenamefont
  {Shi}}]{https://doi.org/10.1029/2011MS000074}%
  \BibitemOpen
  \bibfield  {author} {\bibinfo {author} {\bibfnamefont {Steven~J.}\
  \bibnamefont {Ghan}}, \bibinfo {author} {\bibfnamefont {Hayder}\ \bibnamefont
  {Abdul-Razzak}}, \bibinfo {author} {\bibfnamefont {Athanasios}\ \bibnamefont
  {Nenes}}, \bibinfo {author} {\bibfnamefont {Yi}~\bibnamefont {Ming}},
  \bibinfo {author} {\bibfnamefont {Xiaohong}\ \bibnamefont {Liu}}, \bibinfo
  {author} {\bibfnamefont {Mikhail}\ \bibnamefont {Ovchinnikov}}, \bibinfo
  {author} {\bibfnamefont {Ben}\ \bibnamefont {Shipway}}, \bibinfo {author}
  {\bibfnamefont {Nicholas}\ \bibnamefont {Meskhidze}}, \bibinfo {author}
  {\bibfnamefont {Jun}\ \bibnamefont {Xu}}, \ and\ \bibinfo {author}
  {\bibfnamefont {Xiangjun}\ \bibnamefont {Shi}},\ }\bibfield  {title}
  {\enquote {\bibinfo {title} {Droplet nucleation: Physically-based
  parameterizations and comparative evaluation},}\ }\href {\doibase
  https://doi.org/10.1029/2011MS000074} {\bibfield  {journal} {\bibinfo
  {journal} {Journal of Advances in Modeling Earth Systems}\ }\textbf {\bibinfo
  {volume} {3}} (\bibinfo {year} {2011}),\
  https://doi.org/10.1029/2011MS000074},\ \Eprint
  {http://arxiv.org/abs/https://agupubs.onlinelibrary.wiley.com/doi/pdf/10.1029/2011MS000074}
  {https://agupubs.onlinelibrary.wiley.com/doi/pdf/10.1029/2011MS000074}
  \BibitemShut {NoStop}%
\bibitem [{\citenamefont {Gallo}\ \emph {et~al.}(2021)\citenamefont {Gallo},
  \citenamefont {Magaletti},\ and\ \citenamefont {Casciola}}]{Gallo2021}%
  \BibitemOpen
  \bibfield  {author} {\bibinfo {author} {\bibfnamefont {Mirko}\ \bibnamefont
  {Gallo}}, \bibinfo {author} {\bibfnamefont {Francesco}\ \bibnamefont
  {Magaletti}}, \ and\ \bibinfo {author} {\bibfnamefont {Carlo~Massimo}\
  \bibnamefont {Casciola}},\ }\bibfield  {title} {\enquote {\bibinfo {title}
  {{Heterogeneous bubble nucleation dynamics}},}\ }\href {\doibase
  10.1017/jfm.2020.761} {\bibfield  {journal} {\bibinfo  {journal} {Journal of
  Fluid Mechanics}\ }\textbf {\bibinfo {volume} {906}},\ \bibinfo {pages} {A20}
  (\bibinfo {year} {2021})}\BibitemShut {NoStop}%
\bibitem [{\citenamefont {Moss}(1985)}]{PhysRevD.32.1333}%
  \BibitemOpen
  \bibfield  {author} {\bibinfo {author} {\bibfnamefont {I.~G.}\ \bibnamefont
  {Moss}},\ }\bibfield  {title} {\enquote {\bibinfo {title} {Black-hole
  bubbles},}\ }\href {\doibase 10.1103/PhysRevD.32.1333} {\bibfield  {journal}
  {\bibinfo  {journal} {Phys. Rev. D}\ }\textbf {\bibinfo {volume} {32}},\
  \bibinfo {pages} {1333--1344} (\bibinfo {year} {1985})}\BibitemShut {NoStop}%
\bibitem [{\citenamefont {Gregory}\ \emph {et~al.}(2014)\citenamefont
  {Gregory}, \citenamefont {Moss},\ and\ \citenamefont
  {Withers}}]{Gregory_2014}%
  \BibitemOpen
  \bibfield  {author} {\bibinfo {author} {\bibfnamefont {Ruth}\ \bibnamefont
  {Gregory}}, \bibinfo {author} {\bibfnamefont {Ian~G.}\ \bibnamefont {Moss}},
  \ and\ \bibinfo {author} {\bibfnamefont {Benjamin}\ \bibnamefont {Withers}},\
  }\bibfield  {title} {\enquote {\bibinfo {title} {Black holes as bubble
  nucleation sites},}\ }\href {\doibase 10.1007/jhep03(2014)081} {\bibfield
  {journal} {\bibinfo  {journal} {Journal of High Energy Physics}\ }\textbf
  {\bibinfo {volume} {2014}} (\bibinfo {year} {2014}),\
  10.1007/jhep03(2014)081}\BibitemShut {NoStop}%
\bibitem [{\citenamefont {Burda}\ \emph
  {et~al.}(2015{\natexlab{a}})\citenamefont {Burda}, \citenamefont {Gregory},\
  and\ \citenamefont {Moss}}]{Burda_2015}%
  \BibitemOpen
  \bibfield  {author} {\bibinfo {author} {\bibfnamefont {Philipp}\ \bibnamefont
  {Burda}}, \bibinfo {author} {\bibfnamefont {Ruth}\ \bibnamefont {Gregory}}, \
  and\ \bibinfo {author} {\bibfnamefont {Ian~G.}\ \bibnamefont {Moss}},\
  }\bibfield  {title} {\enquote {\bibinfo {title} {Gravity and the stability of
  the higgs vacuum},}\ }\href {\doibase 10.1103/physrevlett.115.071303}
  {\bibfield  {journal} {\bibinfo  {journal} {Physical Review Letters}\
  }\textbf {\bibinfo {volume} {115}} (\bibinfo {year} {2015}{\natexlab{a}}),\
  10.1103/physrevlett.115.071303}\BibitemShut {NoStop}%
\bibitem [{\citenamefont {Burda}\ \emph
  {et~al.}(2015{\natexlab{b}})\citenamefont {Burda}, \citenamefont {Gregory},\
  and\ \citenamefont {Moss}}]{Burda_2015B}%
  \BibitemOpen
  \bibfield  {author} {\bibinfo {author} {\bibfnamefont {Philipp}\ \bibnamefont
  {Burda}}, \bibinfo {author} {\bibfnamefont {Ruth}\ \bibnamefont {Gregory}}, \
  and\ \bibinfo {author} {\bibfnamefont {Ian~G.}\ \bibnamefont {Moss}},\
  }\bibfield  {title} {\enquote {\bibinfo {title} {Vacuum metastability with
  black holes},}\ }\href {\doibase 10.1007/jhep08(2015)114} {\bibfield
  {journal} {\bibinfo  {journal} {Journal of High Energy Physics}\ }\textbf
  {\bibinfo {volume} {2015}} (\bibinfo {year} {2015}{\natexlab{b}}),\
  10.1007/jhep08(2015)114}\BibitemShut {NoStop}%
\bibitem [{\citenamefont {Burda}\ \emph {et~al.}(2016)\citenamefont {Burda},
  \citenamefont {Gregory},\ and\ \citenamefont {Moss}}]{Burda_2016}%
  \BibitemOpen
  \bibfield  {author} {\bibinfo {author} {\bibfnamefont {Philipp}\ \bibnamefont
  {Burda}}, \bibinfo {author} {\bibfnamefont {Ruth}\ \bibnamefont {Gregory}}, \
  and\ \bibinfo {author} {\bibfnamefont {Ian~G.}\ \bibnamefont {Moss}},\
  }\bibfield  {title} {\enquote {\bibinfo {title} {The fate of the higgs
  vacuum},}\ }\href {\doibase 10.1007/jhep06(2016)025} {\bibfield  {journal}
  {\bibinfo  {journal} {Journal of High Energy Physics}\ }\textbf {\bibinfo
  {volume} {2016}} (\bibinfo {year} {2016}),\
  10.1007/jhep06(2016)025}\BibitemShut {NoStop}%
\bibitem [{\citenamefont {Billam}\ \emph
  {et~al.}(2019{\natexlab{a}})\citenamefont {Billam}, \citenamefont {Gregory},
  \citenamefont {Michel},\ and\ \citenamefont {Moss}}]{Billam:2018pvp}%
  \BibitemOpen
  \bibfield  {author} {\bibinfo {author} {\bibfnamefont {Thomas~P.}\
  \bibnamefont {Billam}}, \bibinfo {author} {\bibfnamefont {Ruth}\ \bibnamefont
  {Gregory}}, \bibinfo {author} {\bibfnamefont {Florent}\ \bibnamefont
  {Michel}}, \ and\ \bibinfo {author} {\bibfnamefont {Ian~G.}\ \bibnamefont
  {Moss}},\ }\bibfield  {title} {\enquote {\bibinfo {title} {{Simulating seeded
  vacuum decay in a cold atom system}},}\ }\href {\doibase
  10.1103/PhysRevD.100.065016} {\bibfield  {journal} {\bibinfo  {journal}
  {Phys. Rev. D}\ }\textbf {\bibinfo {volume} {100}},\ \bibinfo {pages}
  {065016} (\bibinfo {year} {2019}{\natexlab{a}})},\ \Eprint
  {http://arxiv.org/abs/1811.09169} {arXiv:1811.09169 [hep-th]} \BibitemShut
  {NoStop}%
\bibitem [{\citenamefont {Czech}(2012)}]{Czech_2012}%
  \BibitemOpen
  \bibfield  {author} {\bibinfo {author} {\bibfnamefont {Bartłomiej}\
  \bibnamefont {Czech}},\ }\bibfield  {title} {\enquote {\bibinfo {title} {A
  novel channel for vacuum decay},}\ }\href {\doibase
  10.1016/j.physletb.2012.06.018} {\bibfield  {journal} {\bibinfo  {journal}
  {Physics Letters B}\ }\textbf {\bibinfo {volume} {713}},\ \bibinfo {pages}
  {331–334} (\bibinfo {year} {2012})}\BibitemShut {NoStop}%
\bibitem [{\citenamefont {Affleck}\ and\ \citenamefont
  {De~Luccia}(1979)}]{PhysRevD.20.3168}%
  \BibitemOpen
  \bibfield  {author} {\bibinfo {author} {\bibfnamefont {Ian~K.}\ \bibnamefont
  {Affleck}}\ and\ \bibinfo {author} {\bibfnamefont {Frank}\ \bibnamefont
  {De~Luccia}},\ }\bibfield  {title} {\enquote {\bibinfo {title} {Induced
  vacuum decay},}\ }\href {\doibase 10.1103/PhysRevD.20.3168} {\bibfield
  {journal} {\bibinfo  {journal} {Phys. Rev. D}\ }\textbf {\bibinfo {volume}
  {20}},\ \bibinfo {pages} {3168--3178} (\bibinfo {year} {1979})}\BibitemShut
  {NoStop}%
\bibitem [{\citenamefont {Selivanov}\ and\ \citenamefont
  {Voloshin}(1985)}]{Selivanov:1985vt}%
  \BibitemOpen
  \bibfield  {author} {\bibinfo {author} {\bibfnamefont {K.~B.}\ \bibnamefont
  {Selivanov}}\ and\ \bibinfo {author} {\bibfnamefont {M.~B.}\ \bibnamefont
  {Voloshin}},\ }\bibfield  {title} {\enquote {\bibinfo {title} {{Destruction
  of false vacuum by massive particles}},}\ }\href@noop {} {\bibfield
  {journal} {\bibinfo  {journal} {JETP Lett.}\ }\textbf {\bibinfo {volume}
  {42}},\ \bibinfo {pages} {422} (\bibinfo {year} {1985})}\BibitemShut
  {NoStop}%
\bibitem [{\citenamefont {Ellis}\ \emph {et~al.}(1990)\citenamefont {Ellis},
  \citenamefont {Linde},\ and\ \citenamefont {Sher}}]{Ellis:1990bv}%
  \BibitemOpen
  \bibfield  {author} {\bibinfo {author} {\bibfnamefont {John~R.}\ \bibnamefont
  {Ellis}}, \bibinfo {author} {\bibfnamefont {Andrei~D.}\ \bibnamefont
  {Linde}}, \ and\ \bibinfo {author} {\bibfnamefont {Marc}\ \bibnamefont
  {Sher}},\ }\bibfield  {title} {\enquote {\bibinfo {title} {{Vacuum stability,
  wormholes, cosmic rays and the cosmological bounds on m(t) and m(H)}},}\
  }\href {\doibase 10.1016/0370-2693(90)90862-Z} {\bibfield  {journal}
  {\bibinfo  {journal} {Phys. Lett. B}\ }\textbf {\bibinfo {volume} {252}},\
  \bibinfo {pages} {203--211} (\bibinfo {year} {1990})}\BibitemShut {NoStop}%
\bibitem [{\citenamefont {Enqvist}\ and\ \citenamefont
  {McDonald}(1998)}]{Enqvist:1997wv}%
  \BibitemOpen
  \bibfield  {author} {\bibinfo {author} {\bibfnamefont {Kari}\ \bibnamefont
  {Enqvist}}\ and\ \bibinfo {author} {\bibfnamefont {John}\ \bibnamefont
  {McDonald}},\ }\bibfield  {title} {\enquote {\bibinfo {title} {{Can cosmic
  ray catalysed vacuum decay dominate over tunneling?}}}\ }\href {\doibase
  10.1016/S0550-3213(97)00707-4} {\bibfield  {journal} {\bibinfo  {journal}
  {Nucl. Phys. B}\ }\textbf {\bibinfo {volume} {513}},\ \bibinfo {pages}
  {661--678} (\bibinfo {year} {1998})},\ \Eprint
  {http://arxiv.org/abs/hep-ph/9704431} {arXiv:hep-ph/9704431} \BibitemShut
  {NoStop}%
\bibitem [{\citenamefont {Eral}\ \emph {et~al.}(2011)\citenamefont {Eral},
  \citenamefont {Manukyan},\ and\ \citenamefont {Oh}}]{doi:10.1021/la104628q}%
  \BibitemOpen
  \bibfield  {author} {\bibinfo {author} {\bibfnamefont {H.~B.}\ \bibnamefont
  {Eral}}, \bibinfo {author} {\bibfnamefont {G.}~\bibnamefont {Manukyan}}, \
  and\ \bibinfo {author} {\bibfnamefont {J.~M.}\ \bibnamefont {Oh}},\
  }\bibfield  {title} {\enquote {\bibinfo {title} {Wetting of a drop on a
  sphere},}\ }\href {\doibase 10.1021/la104628q} {\bibfield  {journal}
  {\bibinfo  {journal} {Langmuir}\ }\textbf {\bibinfo {volume} {27}},\ \bibinfo
  {pages} {5340--5346} (\bibinfo {year} {2011})},\ \bibinfo {note} {pMID:
  21466229},\ \Eprint {http://arxiv.org/abs/https://doi.org/10.1021/la104628q}
  {https://doi.org/10.1021/la104628q} \BibitemShut {NoStop}%
\bibitem [{\citenamefont {Zenesini}\ \emph {et~al.}(2024)\citenamefont
  {Zenesini}, \citenamefont {Berti}, \citenamefont {Cominotti}, \citenamefont
  {Rogora}, \citenamefont {Moss}, \citenamefont {Billam}, \citenamefont
  {Carusotto}, \citenamefont {Lamporesi}, \citenamefont {Recati},\ and\
  \citenamefont {Ferrari}}]{Zenesini:2023afv}%
  \BibitemOpen
  \bibfield  {author} {\bibinfo {author} {\bibfnamefont {Alessandro}\
  \bibnamefont {Zenesini}}, \bibinfo {author} {\bibfnamefont {Anna}\
  \bibnamefont {Berti}}, \bibinfo {author} {\bibfnamefont {Riccardo}\
  \bibnamefont {Cominotti}}, \bibinfo {author} {\bibfnamefont {Chiara}\
  \bibnamefont {Rogora}}, \bibinfo {author} {\bibfnamefont {Ian~G.}\
  \bibnamefont {Moss}}, \bibinfo {author} {\bibfnamefont {Thomas~P.}\
  \bibnamefont {Billam}}, \bibinfo {author} {\bibfnamefont {Iacopo}\
  \bibnamefont {Carusotto}}, \bibinfo {author} {\bibfnamefont {Giacomo}\
  \bibnamefont {Lamporesi}}, \bibinfo {author} {\bibfnamefont {Alessio}\
  \bibnamefont {Recati}}, \ and\ \bibinfo {author} {\bibfnamefont {Gabriele}\
  \bibnamefont {Ferrari}},\ }\bibfield  {title} {\enquote {\bibinfo {title}
  {{False vacuum decay via bubble formation in ferromagnetic superfluids}},}\
  }\href {\doibase 10.1038/s41567-023-02345-4} {\bibfield  {journal} {\bibinfo
  {journal} {Nature Phys.}\ }\textbf {\bibinfo {volume} {20}},\ \bibinfo
  {pages} {558--563} (\bibinfo {year} {2024})},\ \Eprint
  {http://arxiv.org/abs/2305.05225} {arXiv:2305.05225 [hep-ph]} \BibitemShut
  {NoStop}%
\bibitem [{\citenamefont {{Fialko}}\ \emph {et~al.}(2015)\citenamefont
  {{Fialko}}, \citenamefont {{Opanchuk}}, \citenamefont {{Sidorov}},
  \citenamefont {{Drummond}},\ and\ \citenamefont {{Brand}}}]{FialkoFate2015}%
  \BibitemOpen
  \bibfield  {author} {\bibinfo {author} {\bibfnamefont {O.}~\bibnamefont
  {{Fialko}}}, \bibinfo {author} {\bibfnamefont {B.}~\bibnamefont
  {{Opanchuk}}}, \bibinfo {author} {\bibfnamefont {A.~I.}\ \bibnamefont
  {{Sidorov}}}, \bibinfo {author} {\bibfnamefont {P.~D.}\ \bibnamefont
  {{Drummond}}}, \ and\ \bibinfo {author} {\bibfnamefont {J.}~\bibnamefont
  {{Brand}}},\ }\bibfield  {title} {\enquote {\bibinfo {title} {{Fate of the
  false vacuum: Towards realization with ultra-cold atoms}},}\ }\href {\doibase
  10.1209/0295-5075/110/56001} {\bibfield  {journal} {\bibinfo  {journal} {EPL
  (Europhysics Letters)}\ }\textbf {\bibinfo {volume} {110}},\ \bibinfo {eid}
  {56001} (\bibinfo {year} {2015})},\ \Eprint {http://arxiv.org/abs/1408.1163}
  {arXiv:1408.1163 [cond-mat.quant-gas]} \BibitemShut {NoStop}%
\bibitem [{\citenamefont {{Fialko}}\ \emph {et~al.}(2017)\citenamefont
  {{Fialko}}, \citenamefont {{Opanchuk}}, \citenamefont {{Sidorov}},
  \citenamefont {{Drummond}},\ and\ \citenamefont
  {{Brand}}}]{FialkoUniverse2017}%
  \BibitemOpen
  \bibfield  {author} {\bibinfo {author} {\bibfnamefont {O.}~\bibnamefont
  {{Fialko}}}, \bibinfo {author} {\bibfnamefont {B.}~\bibnamefont
  {{Opanchuk}}}, \bibinfo {author} {\bibfnamefont {A.~I.}\ \bibnamefont
  {{Sidorov}}}, \bibinfo {author} {\bibfnamefont {P.~D.}\ \bibnamefont
  {{Drummond}}}, \ and\ \bibinfo {author} {\bibfnamefont {J.}~\bibnamefont
  {{Brand}}},\ }\bibfield  {title} {\enquote {\bibinfo {title} {{The universe
  on a table top: engineering quantum decay of a relativistic scalar field from
  a metastable vacuum}},}\ }\href {\doibase 10.1088/1361-6455/50/2/024003}
  {\bibfield  {journal} {\bibinfo  {journal} {Journal of Physics B Atomic
  Molecular Physics}\ }\textbf {\bibinfo {volume} {50}},\ \bibinfo {eid}
  {024003} (\bibinfo {year} {2017})},\ \Eprint
  {http://arxiv.org/abs/1607.01460} {arXiv:1607.01460 [cond-mat.quant-gas]}
  \BibitemShut {NoStop}%
\bibitem [{\citenamefont {Billam}\ \emph {et~al.}(2023)\citenamefont {Billam},
  \citenamefont {Brown},\ and\ \citenamefont {Moss}}]{Billam:2022ykl}%
  \BibitemOpen
  \bibfield  {author} {\bibinfo {author} {\bibfnamefont {Thomas~P.}\
  \bibnamefont {Billam}}, \bibinfo {author} {\bibfnamefont {Kate}\ \bibnamefont
  {Brown}}, \ and\ \bibinfo {author} {\bibfnamefont {Ian~G.}\ \bibnamefont
  {Moss}},\ }\bibfield  {title} {\enquote {\bibinfo {title} {{Bubble nucleation
  in a cold spin 1 gas}},}\ }\href {\doibase 10.1088/1367-2630/accca2}
  {\bibfield  {journal} {\bibinfo  {journal} {New J. Phys.}\ }\textbf {\bibinfo
  {volume} {25}},\ \bibinfo {pages} {043028} (\bibinfo {year} {2023})},\
  \Eprint {http://arxiv.org/abs/2212.03621} {arXiv:2212.03621
  [cond-mat.quant-gas]} \BibitemShut {NoStop}%
\bibitem [{\citenamefont {Braden}\ \emph {et~al.}(2019)\citenamefont {Braden},
  \citenamefont {Johnson}, \citenamefont {Peiris}, \citenamefont {Pontzen},\
  and\ \citenamefont {Weinfurtner}}]{Braden:2018tky}%
  \BibitemOpen
  \bibfield  {author} {\bibinfo {author} {\bibfnamefont {Jonathan}\
  \bibnamefont {Braden}}, \bibinfo {author} {\bibfnamefont {Matthew~C.}\
  \bibnamefont {Johnson}}, \bibinfo {author} {\bibfnamefont {Hiranya~V.}\
  \bibnamefont {Peiris}}, \bibinfo {author} {\bibfnamefont {Andrew}\
  \bibnamefont {Pontzen}}, \ and\ \bibinfo {author} {\bibfnamefont {Silke}\
  \bibnamefont {Weinfurtner}},\ }\bibfield  {title} {\enquote {\bibinfo {title}
  {{New Semiclassical Picture of Vacuum Decay}},}\ }\href {\doibase
  10.1103/PhysRevLett.123.031601} {\bibfield  {journal} {\bibinfo  {journal}
  {Phys. Rev. Lett.}\ }\textbf {\bibinfo {volume} {123}},\ \bibinfo {pages}
  {031601} (\bibinfo {year} {2019})},\ \Eprint
  {http://arxiv.org/abs/1806.06069} {arXiv:1806.06069 [hep-th]} \BibitemShut
  {NoStop}%
\bibitem [{\citenamefont {Billam}\ \emph
  {et~al.}(2019{\natexlab{b}})\citenamefont {Billam}, \citenamefont {Gregory},
  \citenamefont {Michel},\ and\ \citenamefont {Moss}}]{BillamSimulating2019}%
  \BibitemOpen
  \bibfield  {author} {\bibinfo {author} {\bibfnamefont {Thomas~P.}\
  \bibnamefont {Billam}}, \bibinfo {author} {\bibfnamefont {Ruth}\ \bibnamefont
  {Gregory}}, \bibinfo {author} {\bibfnamefont {Florent}\ \bibnamefont
  {Michel}}, \ and\ \bibinfo {author} {\bibfnamefont {Ian~G.}\ \bibnamefont
  {Moss}},\ }\bibfield  {title} {\enquote {\bibinfo {title} {Simulating seeded
  vacuum decay in a cold atom system},}\ }\href {\doibase
  10.1103/PhysRevD.100.065016} {\bibfield  {journal} {\bibinfo  {journal}
  {Phys. Rev. D}\ }\textbf {\bibinfo {volume} {100}},\ \bibinfo {pages}
  {065016} (\bibinfo {year} {2019}{\natexlab{b}})},\ \Eprint
  {http://arxiv.org/abs/1811.09169} {arXiv:1811.09169 [hep-th]} \BibitemShut
  {NoStop}%
\bibitem [{\citenamefont {Jenkins}\ \emph {et~al.}(2024)\citenamefont
  {Jenkins}, \citenamefont {Braden}, \citenamefont {Peiris}, \citenamefont
  {Pontzen}, \citenamefont {Johnson},\ and\ \citenamefont
  {Weinfurtner}}]{Jenkins:2023eez}%
  \BibitemOpen
  \bibfield  {author} {\bibinfo {author} {\bibfnamefont {Alexander~C.}\
  \bibnamefont {Jenkins}}, \bibinfo {author} {\bibfnamefont {Jonathan}\
  \bibnamefont {Braden}}, \bibinfo {author} {\bibfnamefont {Hiranya~V.}\
  \bibnamefont {Peiris}}, \bibinfo {author} {\bibfnamefont {Andrew}\
  \bibnamefont {Pontzen}}, \bibinfo {author} {\bibfnamefont {Matthew~C.}\
  \bibnamefont {Johnson}}, \ and\ \bibinfo {author} {\bibfnamefont {Silke}\
  \bibnamefont {Weinfurtner}},\ }\bibfield  {title} {\enquote {\bibinfo {title}
  {{Analog vacuum decay from vacuum initial conditions}},}\ }\href {\doibase
  10.1103/PhysRevD.109.023506} {\bibfield  {journal} {\bibinfo  {journal}
  {Phys. Rev. D}\ }\textbf {\bibinfo {volume} {109}},\ \bibinfo {pages}
  {023506} (\bibinfo {year} {2024})},\ \Eprint
  {http://arxiv.org/abs/2307.02549} {arXiv:2307.02549 [cond-mat.quant-gas]}
  \BibitemShut {NoStop}%
\bibitem [{\citenamefont {Billam}\ \emph {et~al.}(2020)\citenamefont {Billam},
  \citenamefont {Brown},\ and\ \citenamefont {Moss}}]{Billam:2020xna}%
  \BibitemOpen
  \bibfield  {author} {\bibinfo {author} {\bibfnamefont {Thomas~P.}\
  \bibnamefont {Billam}}, \bibinfo {author} {\bibfnamefont {Kate}\ \bibnamefont
  {Brown}}, \ and\ \bibinfo {author} {\bibfnamefont {Ian~G.}\ \bibnamefont
  {Moss}},\ }\bibfield  {title} {\enquote {\bibinfo {title} {{Simulating
  cosmological supercooling with a cold atom system}},}\ }\href {\doibase
  10.1103/PhysRevA.102.043324} {\bibfield  {journal} {\bibinfo  {journal}
  {Phys. Rev. A}\ }\textbf {\bibinfo {volume} {102}},\ \bibinfo {pages}
  {043324} (\bibinfo {year} {2020})},\ \Eprint
  {http://arxiv.org/abs/2006.09820} {arXiv:2006.09820 [cond-mat.quant-gas]}
  \BibitemShut {NoStop}%
\bibitem [{\citenamefont {Billam}\ \emph {et~al.}(2021)\citenamefont {Billam},
  \citenamefont {Brown}, \citenamefont {Groszek},\ and\ \citenamefont
  {Moss}}]{Billam:2021psh}%
  \BibitemOpen
  \bibfield  {author} {\bibinfo {author} {\bibfnamefont {Thomas~P.}\
  \bibnamefont {Billam}}, \bibinfo {author} {\bibfnamefont {Kate}\ \bibnamefont
  {Brown}}, \bibinfo {author} {\bibfnamefont {Andrew~J.}\ \bibnamefont
  {Groszek}}, \ and\ \bibinfo {author} {\bibfnamefont {Ian~G.}\ \bibnamefont
  {Moss}},\ }\bibfield  {title} {\enquote {\bibinfo {title} {{Simulating
  cosmological supercooling with a cold atom system. II. Thermal damping and
  parametric instability}},}\ }\href {\doibase 10.1103/PhysRevA.104.053309}
  {\bibfield  {journal} {\bibinfo  {journal} {Phys. Rev. A}\ }\textbf {\bibinfo
  {volume} {104}},\ \bibinfo {pages} {053309} (\bibinfo {year}
  {2021})}\BibitemShut {NoStop}%
\bibitem [{\citenamefont {P\^\i{}rvu}\ \emph {et~al.}(2023)\citenamefont
  {P\^\i{}rvu}, \citenamefont {Johnson},\ and\ \citenamefont
  {Sibiryakov}}]{Pirvu:2023plk}%
  \BibitemOpen
  \bibfield  {author} {\bibinfo {author} {\bibfnamefont {Dalila}\ \bibnamefont
  {P\^\i{}rvu}}, \bibinfo {author} {\bibfnamefont {Matthew~C.}\ \bibnamefont
  {Johnson}}, \ and\ \bibinfo {author} {\bibfnamefont {Sergey}\ \bibnamefont
  {Sibiryakov}},\ }\bibfield  {title} {\enquote {\bibinfo {title} {{Bubble
  velocities and oscillon precursors in first order phase transitions}},}\
  }\href@noop {} {\  (\bibinfo {year} {2023})},\ \Eprint
  {http://arxiv.org/abs/2312.13364} {arXiv:2312.13364 [hep-th]} \BibitemShut
  {NoStop}%
\bibitem [{\citenamefont {Jenkins}\ \emph {et~al.}(2023)\citenamefont
  {Jenkins}, \citenamefont {Moss}, \citenamefont {Billam}, \citenamefont
  {Hadzibabic}, \citenamefont {Peiris},\ and\ \citenamefont
  {Pontzen}}]{Jenkins:2023npg}%
  \BibitemOpen
  \bibfield  {author} {\bibinfo {author} {\bibfnamefont {Alexander~C.}\
  \bibnamefont {Jenkins}}, \bibinfo {author} {\bibfnamefont {Ian~G.}\
  \bibnamefont {Moss}}, \bibinfo {author} {\bibfnamefont {Thomas~P.}\
  \bibnamefont {Billam}}, \bibinfo {author} {\bibfnamefont {Zoran}\
  \bibnamefont {Hadzibabic}}, \bibinfo {author} {\bibfnamefont {Hiranya~V.}\
  \bibnamefont {Peiris}}, \ and\ \bibinfo {author} {\bibfnamefont {Andrew}\
  \bibnamefont {Pontzen}},\ }\bibfield  {title} {\enquote {\bibinfo {title}
  {{Generalized cold-atom simulators for vacuum decay}},}\ }\href@noop {} {\
  (\bibinfo {year} {2023})},\ \Eprint {http://arxiv.org/abs/2311.02156}
  {arXiv:2311.02156 [cond-mat.quant-gas]} \BibitemShut {NoStop}%
\bibitem [{\citenamefont {Langer}(1969)}]{Langer:1969bc}%
  \BibitemOpen
  \bibfield  {author} {\bibinfo {author} {\bibfnamefont {J.~S.}\ \bibnamefont
  {Langer}},\ }\bibfield  {title} {\enquote {\bibinfo {title} {{Statistical
  theory of the decay of metastable states}},}\ }\href {\doibase
  10.1016/0003-4916(69)90153-5} {\bibfield  {journal} {\bibinfo  {journal}
  {Annals Phys.}\ }\textbf {\bibinfo {volume} {54}},\ \bibinfo {pages}
  {258--275} (\bibinfo {year} {1969})}\BibitemShut {NoStop}%
\bibitem [{\citenamefont {Soleimani}\ \emph {et~al.}(2013)\citenamefont
  {Soleimani}, \citenamefont {Hill},\ and\ \citenamefont {van~de
  Ven}}]{doi:10.1021/la403088r}%
  \BibitemOpen
  \bibfield  {author} {\bibinfo {author} {\bibfnamefont {Majid}\ \bibnamefont
  {Soleimani}}, \bibinfo {author} {\bibfnamefont {Reghan~J.}\ \bibnamefont
  {Hill}}, \ and\ \bibinfo {author} {\bibfnamefont {Theo G.~M.}\ \bibnamefont
  {van~de Ven}},\ }\bibfield  {title} {\enquote {\bibinfo {title} {Bubbles and
  drops on curved surfaces},}\ }\href {\doibase 10.1021/la403088r} {\bibfield
  {journal} {\bibinfo  {journal} {Langmuir}\ }\textbf {\bibinfo {volume}
  {29}},\ \bibinfo {pages} {14168--14177} (\bibinfo {year} {2013})},\ \Eprint
  {http://arxiv.org/abs/https://doi.org/10.1021/la403088r}
  {https://doi.org/10.1021/la403088r} \BibitemShut {NoStop}%
\bibitem [{\citenamefont {Bradley}\ and\ \citenamefont
  {Blakie}(2014)}]{BradleyStochastic2014}%
  \BibitemOpen
  \bibfield  {author} {\bibinfo {author} {\bibfnamefont {Ashton~S.}\
  \bibnamefont {Bradley}}\ and\ \bibinfo {author} {\bibfnamefont {P.~Blair}\
  \bibnamefont {Blakie}},\ }\bibfield  {title} {\enquote {\bibinfo {title}
  {Stochastic projected gross-pitaevskii equation for spinor and multicomponent
  condensates},}\ }\href {\doibase 10.1103/PhysRevA.90.023631} {\bibfield
  {journal} {\bibinfo  {journal} {Phys. Rev. A}\ }\textbf {\bibinfo {volume}
  {90}},\ \bibinfo {pages} {023631} (\bibinfo {year} {2014})},\ \Eprint
  {http://arxiv.org/abs/1406.2029} {arXiv:1406.2029 [cond-mat.quant-gas]}
  \BibitemShut {NoStop}%
\bibitem [{\citenamefont {Canaletti}\ and\ \citenamefont {Moss}()}]{data}%
  \BibitemOpen
  \bibfield  {author} {\bibinfo {author} {\bibfnamefont {Matteo}\ \bibnamefont
  {Canaletti}}\ and\ \bibinfo {author} {\bibfnamefont {Ian~G}\ \bibnamefont
  {Moss}},\ }\bibfield  {title} {\enquote {\bibinfo {title} {Data supporting
  publication: Seeding decay of the false vacuum},}\ }\href {\doibase
  10.25405/data.ncl.26797480} {\ 10.25405/data.ncl.26797480}\BibitemShut
  {NoStop}%
\bibitem [{\citenamefont {Abed}\ and\ \citenamefont
  {Moss}(2023)}]{Abed:2020lcf}%
  \BibitemOpen
  \bibfield  {author} {\bibinfo {author} {\bibfnamefont {Mario~Gutierrez}\
  \bibnamefont {Abed}}\ and\ \bibinfo {author} {\bibfnamefont {Ian~G.}\
  \bibnamefont {Moss}},\ }\bibfield  {title} {\enquote {\bibinfo {title}
  {{Bubble nucleation at zero and nonzero temperatures}},}\ }\href {\doibase
  10.1103/PhysRevD.107.076027} {\bibfield  {journal} {\bibinfo  {journal}
  {Phys. Rev. D}\ }\textbf {\bibinfo {volume} {107}},\ \bibinfo {pages}
  {076027} (\bibinfo {year} {2023})},\ \Eprint
  {http://arxiv.org/abs/2006.06289} {arXiv:2006.06289 [hep-th]} \BibitemShut
  {NoStop}%
\end{thebibliography}%

\end{document}